# Eshelbian mechanics of novel materials: Quasicrystals

Markus Lazar* and Eleni Agiasofitou[†]

*Heisenberg Research Group, Department of Physics*
*Darmstadt University of Technology*
*Hochschulstr. 6, D-64289 Darmstadt, Germany*

**Abstract** In this work, the so-called Eshelbian or configurational mechanics of quasicrystals is presented. Quasicrystals are considered as a prototype of novel materials. Material balance laws for quasicrystalline materials with dislocations are derived in the framework of generalized incompatible elasticity theory of quasicrystals. Translations, scaling transformations as well as rotations are examined; the latter presents particular interest due to the quasicrystalline structure. This derivation provides important quantities of the Eshelbian mechanics, as the Eshelby stress tensor, the scaling flux vector, the angular momentum tensor, the configurational forces (Peach-Koehler force, Cherepanov force, inhomogeneity force or Eshelby force), the configurational work and the configurational vector moments for dislocations in quasicrystals. The corresponding $J$-, $M$-, and $L$-integrals for dislocation loops and straight dislocations in quasicrystals are derived and discussed. Moreover, the explicit formulas of the $J$-, $M$-, and $L$-integrals for parallel screw dislocations in one-dimensional hexagonal quasicrystals are obtained. Through this derivation, the physical interpretation of the $J$-, $M$-, $L$-integrals for dislocations in quasicrystals is revealed and their connection to the Peach-Koehler force, the interaction energy and the rotational vector moment (torque) of dislocations in quasicrystals is established.

## 1. Introduction

Quasicrystals were discovered by Shechtman in 1982 (see [Shechtman *et al.*, 1984]). Quasicrystals belong to aperiodic crystals and possess a long-range orientational order but no translational symmetry. Due to the discovery of quasicrystals, the International Union of Crystallography changed the official definition of a crystal in 1992. Shechtman was awarded the 2011 Nobel Prize in Chemistry for his great discovery bringing quasicrystals to the front of the scientific panel. During the last three decades researchers from various scientific directions exploring the geometric structure and the properties (e.g., mechanical, electronic, thermodynamical, chemical) of quasicrystals have shown, that quasicrystals have some particular properties which could be characterized as desirable properties. Among these properties we mention low friction coefficient [Dubois *et al.*, 1991], high wear resistance [Bloom *et al.*, 2003], and low adhesion [Park *et al.*, 2004]. These properties give quasicrystals advantages in comparison to other conventional materials used today (e.g., [Kenzari *et al.*, 2012]), and can have desirable and promising applications to technology. Quasicrystals represent an interesting and important class of novel materials.

* *E-mail address:* lazar@fkp.tu-darmstadt.de (M. Lazar).
† *E-mail address:* agiasofitou@mechanik.tu-darmstadt.de (E. Agiasofitou).

The basis of the continuum theory of solid quasicrystals is set up by two elementary excitations; the phonons and the phasons [Bak, 1985; Levine *et al.*, 1985]. Quasicrystals can have crystallographic and non-crystallographic point groups. The structure of icosahedral, octagonal, decagonal and dodecagonal quasicrystals have non-crystallographic point-group symmetries. However, a quasicrystal is not necessary to be associated with non-crystallographic point-group symmetry. For instance, a three-dimensional cubic quasicrystal possesses crystallographic point-group symmetry (see [Yang *et al.*, 1993]). To study the elastic properties of quasicrystals, the framework of linear elasticity theory of quasicrystals has been investigated. Levine *et al.* [1985] and Socolar [1989] derived the elastic energies of icosahedral and planar pentagonal, octagonal and dodecagonal quasicrystals. The generalized linear elasticity theory of quasicrystals was first proposed by Ding *et al.* [1993]. Yang *et al.* [1993] investigated the linear elasticity theory of cubic quasicrystals. The generalized linear elasticity theory of one-dimensional quasicrystals was developed by Wang *et al.* [1997].

Quasicrystals are fundamentally anisotropic due to their positional and orientational long-range order. Therefore, the elastic properties are expected to be anisotropic. However, Levine *et al.* [1985] showed that linear phonon elasticity of pentagonal and icosahedral quasicrystals is isotropic, and Socolar [1989] showed that linear phonon elasticity of planar octagonal, decagonal and dodecagonal quasicrystals is isotropic (see also [Ding *et al.*, 1993]). Moreover, it was shown within linear phonon elasticity that many quasicrystals behave essentially like isotropic media [Gong *et al.*, 2006a,b]. The reason is that if the fold of the symmetry axes which quasicrystals have are generally higher than four, then it leads to the $\infty$-fold symmetry axes for the fourth rank phonon elastic tensor. According to the Hermann theorem [Hermann, 1934] the minimal rank $r$ of a tensor which is necessary in order to reveal the anisotropy of a symmetry is related to the order of the symmetry rotation; e.g. for icosahedral quasicrystals, $r = 5$, and for decagonal quasicrystals, $r = 10$ (see also [Ripamonti, 1987]).

Quasicrystals are brittle at room temperature and dislocations are apparent, influencing the macroscopic plastic behavior of these materials (see, e.g., [Feuerbacher, 2012]). Hence, the modeling of defects (cracks and dislocations) is of high importance. It is well-known that in classical elasticity and generalized elasticity theories, the so-called Eshelbian or configurational mechanics offers an appropriate treatment of defects like cracks, dislocations, disclinations, and inclusions (see, e.g., [Maugin, 1993; Kienzler and Herrmann, 2000; Markenscoff and Gupta, 2006; Li and Wang, 2008]). The Eshelbian or configurational mechanics of quasicrystals is a new and promising research field starting with the works of Shi [2007] and Agiasofitou *et al.* [2010]. Shi [2005, 2007] was the first to study conservation laws and the corresponding path-independent integrals for quasicrystals. The static Peach-Koehler force for straight dislocations in one-dimensional hexagonal quasicrystals was first derived by Li and Fan [1999]. In the framework of fracture mechanics, Fan and Mai [2004] (see also [Fan, 2011]) derived a "generalized Eshelby integral", which has the physical meaning of a $J_1$-integral, in terms of a generalized Eshelby stress tensor for quasicrystals. Using the framework of Eshelbian or configurational mechanics, the general expressions for the vectorial $\boldsymbol{J}$-integral, the dynamical Peach-Koehler force, the static Peach-Koehler force and the corresponding Eshelby stress tensor were first derived by Agiasofitou *et al.* [2010] for arbitrary quasicrystals. Lazar and Agiasofitou [2014] have investigated fundamental aspects of generalized elasticity theory and dislocation theory deriving the three-dimensional elastic Green tensor and all the dislocation key-formulas (e.g., generalized Burgers formula, Mura-Willis formula, Peach-Koehler stress formula) including an application to dislocation loops for arbitrary quasicrystals.

An outstanding task in the defect mechanics of quasicrystals is the derivation of the $\boldsymbol{J}$-, $M$-, and $\boldsymbol{L}$-integrals in the framework of Eshelbian mechanics of quasicrystals. In general, the $\boldsymbol{J}$-, $M$-, and $\boldsymbol{L}$-integrals [Budiansky and Rice, 1973; Eshelby, 1975; Kirchner, 1999] are important for the study of dislocations, cracks in fracture mechanics, and for dislocation based fracture mechanics. In the study of dislocations, it is known that the appearance of the Peach-Koehler force prevents the existence of $\boldsymbol{J}$-, $M$-, and $\boldsymbol{L}$-conservation laws (see e.g., Lazar and Kirchner [2007a,b]). Concerning the $\boldsymbol{J}$-, $M$-, and $\boldsymbol{L}$-integrals for quasicrystals, up to now, only the $\boldsymbol{J}$- and $M$-integrals for plane and antiplane problems were given by Shi [2007] in the framework of compatible elasticity of two-dimensional quasicrystals; and for dislocations, the dynamical $\boldsymbol{J}$-integral has been derived by Agiasofitou *et al.* [2010] in the framework of incompatible elasticity for arbitrary quasicrystals.

The paper is organized into six sections. In Sec. 2, we start by setting the stage of generalized elasticity theory of quasicrystals (incompatible and compatible) with the focus on the presence of dislocations inside the quasicrystalline material. Material balance laws for incompatible elasticity of quasicrystals considering additionally body forces and a non-homogeneous material are derived in Sec. 3. Particularly, we investigate translational, rotational, and dilatational balance laws. The balance laws contain important quantities of Eshelbian mechanics like the Eshelby stress tensor, the scaling flux vector, the angular momentum tensor, the configurational forces, the configurational work, the configurational vector moments and give rise to the corresponding $\boldsymbol{J}$-, $M$-, and $\boldsymbol{L}$-integrals for quasicrystals. In Sec. 4, we derive the $\boldsymbol{J}$-, $M$-, and $\boldsymbol{L}$-integrals for straight dislocations in quasicrystals. Moreover, as a representative example, we calculate the explicit form of the $\boldsymbol{J}$-, $M$-, and $\boldsymbol{L}$-integrals for two parallel screw dislocations in one-dimensional hexagonal quasicrystals. The obtained results reveal the physical meaning and the significance of the examined integrals connecting them with the Peach-Koehler force, the interaction energy and the rotational vector moment (torque) of dislocations in quasicrystals. In Sec. 5, we give the $\boldsymbol{J}$-, $M$-, and $\boldsymbol{L}$-integrals of generalized compatible elasticity theory of quasicrystals. Finally, in Sec. 6, we close our presentation by giving the conclusions.

## 2. Generalized incompatible elasticity theory and dislocations in quasicrystals

In this section, we give the basic equations of generalized elasticity theory of quasicrystals, valid for the incompatible and compatible cases. Next, the basic equations of generalized incompatible elasticity theory are specified in the presence of dislocations in a quasicrystalline material.

An $(n-3)$-dimensional quasicrystal can be generated by the projection of an $n$-dimensional periodic structure to the three-dimensional physical space ($n = 4, 5, 6$). The $n$-dimensional hyperspace (hyperlattice) $E^n$ can be decomposed into the direct sum of two orthogonal subspaces,

$$E^n = E_\parallel^3 \oplus E_\perp^{(n-3)}, \tag{1}$$

where $E_\parallel^3$ is the three-dimensional *physical* or *parallel space of the phonon fields* and $E_\perp^{(n-3)}$ is the $(n-3)$-dimensional *perpendicular space of the phason fields*. For $n = 4, 5, 6$, we speak of one-dimensional, two-dimensional, and three-dimensional quasicrystals, respectively. Indices in the parallel space are denoted by small letters $i, j, k$ with $i, j, k = 1, 2, 3$. Indices in the perpendicular space are denoted by capital letters $A, B$ with $A, B = 3$ for one-dimensional quasicrystals (with quasiperiodicity in $z$-direction), $A, B = 1, 2$ for two-dimensional quasicrystals, and $A, B = 1, 2, 3$ for three-dimensional quasicrystals. For one-dimensional quasicrystals there exist 31 point groups, consisting of triclinic, monoclinic, orthorhombic, tetragonal, trigonal and hexagonal systems, and 10 Laue classes [Wang *et al.*, 1997], which are crystallographic point groups. The two-dimensional quasicrystals have 57 point groups, consisting of 31 crystallographic point groups and 26 non-crystallographic point groups (pentagonal, decagonal, octagonal, dodecagonal systems) [Hu *et al.*, 1996]. The three-dimensional quasicrystals have 60 point groups, consisting of 32 crystallographic point groups (i.e., cubic system) and 28 non-crystallographic point groups (i.e., icosahedral system) [Fan, 2011]. For quasicrystals with non-crystallographic symmetries, the *phonon displacement* $u_i^\parallel$ and the *phason displacement* $u_A^\perp$ transform under different irreducible representations of the point group. Therefore, from the group theoretical point of view, it is useful to use two different coordinate systems for the phonon and phason fields, since the phonon field variable $u_i^\parallel$ is a three-dimensional vector in $E_\parallel^3$ and transforms under the vector representation in $E_\parallel^3$, whereas the phason field variable $u_A^\perp$ is an $(n-3)$-dimensional vector in $E_\perp^{(n-3)}$ and transforms under another irreducible representation in $E_\perp^{(n-3)}$ (e.g., Bak [1985];

Gong *et al.* [2006a,b]). It will become evident that the introduction of these two different coordinate systems with different indices greatly facilitates the calculations providing simultaneously clarity in what concerns the different nature of the phonon and phason fields. It should be noted that there exists an exceptional case; that one of three-dimensional cubic quasicrystals which belong to an important class of quasicrystals. For three-dimensional cubic quasicrystals, the phonon and phason displacement fields transform under the same irreducible vector representation which induces some elastic behavior different from "usual" quasicrystals [Yang *et al.*, 1993]. Therefore, for cubic quasicrystals one can interchange the indices and small letters can be used for both indices, namely for the indices that "live" in the parallel space and for the indices that "live" in the perpendicular space ($i = A = 1, 2, 3$). Throughout the text, phonon fields will be denoted by $(\cdot)^{\parallel}$ and phason fields by $(\cdot)^{\perp}$. All quantities (phonon and phason fields) depend on the so-called material space coordinates (or spatial coordinates) $\boldsymbol{x} \in \mathbb{R}^3$. Notice that in the linear theory of quasicrystals the material space coincides with the parallel space.

In the theory of *generalized linear elasticity of quasicrystals*, the *elastic energy density* stored in a quasicrystal (for the unlocked state) can be written as (e.g., Ding *et al.* [1993]; Hu *et al.* [2000]; Agiasofitou *et al.* [2010])

$$W = \frac{1}{2} \sigma_{ij}^{\parallel} \beta_{ij}^{\parallel} + \frac{1}{2} \sigma_{Aj}^{\perp} \beta_{Aj}^{\perp} \,, \tag{2}$$

where $\sigma_{ij}^{\parallel}$ and $\sigma_{Aj}^{\perp}$ are the *phonon and phason stress tensors*, respectively. We note that the phonon stress is symmetric, $\sigma_{ij}^{\parallel} = \sigma_{ji}^{\parallel}$, whereas the phason stress is asymmetric, $\sigma_{Aj}^{\perp} \neq \sigma_{jA}^{\perp}$. $\beta_{ij}^{\parallel}$ is the *elastic phonon distortion tensor* and $\beta_{Aj}^{\perp}$ denotes the *elastic phason distortion tensor*. It becomes now evident that due to the two different indices, $\sigma_{Aj}^{\perp}$ and $\beta_{Aj}^{\perp}$ are two-point tensors. The explicit form of Eq. (2) reads

$$W = \frac{1}{2} C_{ijkl} \beta_{ij}^{\parallel} \beta_{kl}^{\parallel} + D_{ijBl} \beta_{ij}^{\parallel} \beta_{Bl}^{\perp} + \frac{1}{2} E_{AjBl} \beta_{Aj}^{\perp} \beta_{Bl}^{\perp} \,. \tag{3}$$

Here, the tensors of the elastic moduli of quasicrystals possess the symmetries

$$C_{ijkl} = C_{klij} = C_{ijlk} = C_{jikl} \,, \qquad D_{ijBl} = D_{jiBl} \,, \qquad E_{AjBl} = E_{BlAj} \,, \tag{4}$$

where $C_{ijkl}$ is the tensor of the elastic moduli of phonons, $E_{AjBl}$ is the tensor of the elastic moduli of phasons, and $D_{ijBl}$ is the tensor of the elastic moduli of the phonon-phason coupling. It is worth noting that the fourth rank tensors $E_{AjBl}$ and $D_{ijBl}$ are two-point tensors.

The *equilibrium equations of quasicrystals* read (e.g., Ding *et al.* [1993]; Agiasofitou *et al.* [2010])

$$\sigma_{ij,j}^{\parallel} + f_i^{\parallel} = 0 \,, \tag{5}$$

$$\sigma_{Aj,j}^{\perp} + f_A^{\perp} = 0 \,, \tag{6}$$

where $f_i^{\parallel}$ is the *conventional (phonon) body force density* and $f_A^{\perp}$ is a *generalized (phason) body force density*. The subscript comma denotes partial differentiation $\partial_j = \partial/\partial x_j$ with respect to the spatial coordinates $x_j$. From the energy density (3), we obtain the constitutive relations

$$\sigma_{ij}^{\parallel} = \frac{\partial W}{\partial \beta_{ij}^{\parallel}} = C_{ijkl}\beta_{kl}^{\parallel} + D_{ijBl}\beta_{Bl}^{\perp}, \tag{7}$$

$$\sigma_{Aj}^{\perp} = \frac{\partial W}{\partial \beta_{Aj}^{\perp}} = D_{klAj}\beta_{kl}^{\parallel} + E_{AjBl}\beta_{Bl}^{\perp}. \tag{8}$$

Eqs. (2)–(8) are valid in the framework of generalized incompatible and compatible elasticity theories of quasicrystals.

If dislocations are present inside a quasicrystalline material, then the appropriate framework for dislocations is the theory of generalized incompatible elasticity of quasicrystals [Agiasofitou *et al.*, 2010; Lazar and Agiasofitou, 2014; Ding *et al.*, 1995]. The gradient of the phonon displacement vector $u_i^{\parallel}$ and of the phason displacement vector $u_A^{\perp}$ can be decomposed into their elastic and plastic parts according to (e.g., Agiasofitou *et al.* [2010]; Lazar and Agiasofitou [2014])

$$u_{i,j}^{\parallel} = \beta_{ij}^{\parallel} + \beta_{ij}^{\parallel P}, \qquad u_{A,j}^{\perp} = \beta_{Aj}^{\perp} + \beta_{Aj}^{\perp P}, \tag{9}$$

where $\beta_{ij}^{\parallel P}$ is the *phonon plastic distortion tensor* and $\beta_{Aj}^{\perp P}$ is the *phason plastic distortion tensor*. The *phonon and phason dislocation density tensors* $\alpha_{ij}^{\parallel}$ and $\alpha_{Aj}^{\perp}$, respectively, are defined in terms of the plastic distortion tensors

$$\alpha_{ij}^{\parallel} = -\epsilon_{jkl}\beta_{il,k}^{\parallel P}, \qquad \alpha_{Aj}^{\perp} = -\epsilon_{jkl}\beta_{Al,k}^{\perp P} \tag{10}$$

or in terms of the elastic distortion tensors

$$\alpha_{ij}^{\parallel} = \epsilon_{jkl}\beta_{il,k}^{\parallel}, \qquad \alpha_{Aj}^{\perp} = \epsilon_{jkl}\beta_{Al,k}^{\perp}, \tag{11}$$

where $\epsilon_{jkl}$ denotes the three-dimensional Levi-Civita tensor. Eq. (10) represents the basic definition of the dislocation density tensors of quasicrystals, whereas Eq. (11) is a consequence that the quasicrystal remains perfectly connected under the incompatible distortions. In our notation, the first index of the dislocation density tensor (10) shows the orientation of the Burgers vectors and the second index shows the direction of the dislocation line. In addition, the dislocation density tensors satisfy the following *dislocation Bianchi identities*

$$\alpha_{ij,j}^{\parallel} = 0, \qquad \alpha_{Aj,j}^{\perp} = 0, \tag{12}$$

which mean that dislocations cannot end inside the quasicrystalline medium. It is evident that $\beta_{Aj}^{\perp P}$ and $\alpha_{Aj}^{\perp}$ are also two-point tensors.

Substituting Eqs. (7)–(9) into the equilibrium equations (5) and (6), the *field equations for the phonon and phason displacement fields* are given by the following coupled inhomogeneous Navier equations [Ding *et al.*, 1995; Lazar and Agiasofitou, 2014; Agiasofitou and Lazar, 2014]

$$C_{ijkl}u_{k,lj}^{\parallel} + D_{ijBl}u_{B,lj}^{\perp} = C_{ijkl}\beta_{kl,j}^{\parallel P} + D_{ijBl}\beta_{Bl,j}^{\perp P} - f_i^{\parallel}, \tag{13}$$

$$D_{klAj}u_{k,lj}^{\parallel} + E_{AjBl}u_{B,lj}^{\perp} = D_{klAj}\beta_{kl,j}^{\parallel P} + E_{AjBl}\beta_{Bl,j}^{\perp P} - f_A^{\perp}. \tag{14}$$

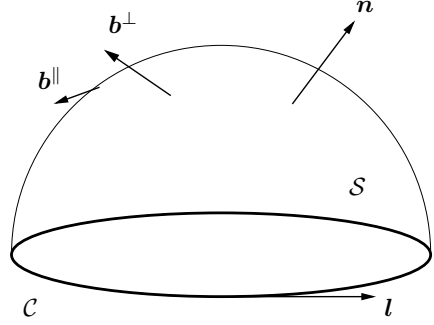

Figure 1. Dislocation loop in quasicrystals.

If we compare Eqs. (13) and (14) with Eqs. (5) and (6), we may introduce *fictitious body force densities caused by the gradient of the plastic distortions* (see, deWit [1973]; Ding *et al.* [1995]; Lazar and Agiasofitou [2014]), namely a *fictitious phonon body force density*

$$f_i^{\parallel \mathrm{P}} = -C_{ijkl}\beta_{kl,j}^{\parallel \mathrm{P}} - D_{ijBl}\beta_{Bl,j}^{\perp \mathrm{P}}$$  (15)

and a *fictitious phason body force density*

$$f_A^{\perp \mathrm{P}} = -D_{klAj}\beta_{kl,j}^{\parallel \mathrm{P}} - E_{AjBl}\beta_{Bl,j}^{\perp \mathrm{P}}\,,$$  (16)

which are non-zero due to the non-uniform (non-constant) plastic fields (or eigendistortions) caused by dislocations. Thus, Eq. (16) gives an example of a phason force density caused by the gradient of the plastic fields of dislocations.

A dislocation in a quasicrystal can be considered as a *hyperdislocation* in the hyperlattice by means of a generalized Volterra process, since the hyperlattice is periodic (see, e.g., Wang and Hu [2002]; Feuerbacher [2012]; Lazar and Agiasofitou [2014]). The Burgers vector $\boldsymbol{b} = (b_i^{\parallel}, b_A^{\perp})$ of the hyperdislocation consists of phonon and phason components which are given by the following surface and line integrals

$$b_i^{\parallel} = \int_\sigma \alpha_{ij}^{\parallel}\,\mathrm{d}S_j = \oint_\gamma \beta_{ij}^{\parallel}\,\mathrm{d}l_j = -\oint_\gamma \beta_{ij}^{\parallel \mathrm{P}}\,\mathrm{d}l_j\,,$$  (17)

$$b_A^{\perp} = \int_\sigma \alpha_{Aj}^{\perp}\,\mathrm{d}S_j = \oint_\gamma \beta_{Aj}^{\perp}\,\mathrm{d}l_j = -\oint_\gamma \beta_{Aj}^{\perp \mathrm{P}}\,\mathrm{d}l_j\,,$$  (18)

where $\gamma$ is the Burgers circuit, which is a closed curve around the dislocation line $\mathcal{C}$, and $\sigma$ is the Burgers surface, bounded by $\gamma$.

A dislocation loop $\mathcal{C}$ is defined as the boundary of a surface (dislocation surface) $\mathcal{S}$ where the phonon displacement has a jump $b_i^{\parallel}$, and the phason displacement has a jump $b_A^{\perp}$ (see Fig. 1). For such a Volterra-type dislocation loop $\mathcal{C}$ in a quasicrystal, the phonon plastic distortion tensor, the phason plastic distortion tensor, the phonon dislocation density tensor and the phason dislocation density tensor are given by (e.g., Agiasofitou *et al.* [2010]; Agiasofitou and Lazar [2014]; Lazar and Agiasofitou [2014])

$$\beta_{ij}^{\parallel P} = -b_i^{\parallel}\delta_j(\mathcal{S}) = -b_i^{\parallel}\int_{\mathcal{S}}\delta(\boldsymbol{x}-\boldsymbol{x}')\,\mathrm{d}S_j' \,, \tag{19}$$

$$\beta_{Aj}^{\perp P} = -b_A^{\perp}\delta_j(\mathcal{S}) = -b_A^{\perp}\int_{\mathcal{S}}\delta(\boldsymbol{x}-\boldsymbol{x}')\,\mathrm{d}S_j' \,, \tag{20}$$

$$\alpha_{ij}^{\parallel} = b_i^{\parallel}\delta_j(\mathcal{C}) = b_i^{\parallel}\oint_{\mathcal{C}}\delta(\boldsymbol{x}-\boldsymbol{x}')\,\mathrm{d}l_j' \,, \tag{21}$$

$$\alpha_{Aj}^{\perp} = b_A^{\perp}\delta_j(\mathcal{C}) = b_A^{\perp}\oint_{\mathcal{C}}\delta(\boldsymbol{x}-\boldsymbol{x}')\,\mathrm{d}l_j' \,, \tag{22}$$

where $\delta_j(\mathcal{S})$ and $\delta_j(\mathcal{C})$ are the Dirac delta functions on the surface $\mathcal{S}$ and the curve $\mathcal{C}$, respectively, $\mathrm{d}l_j'$ denotes the dislocation line element at $\boldsymbol{x}'$ and $\mathrm{d}S_j'$ is the corresponding dislocation loop area.

## 3. Material balance laws for quasicrystals

In this section, we construct material balance laws for a quasicrystalline material with dislocations following the procedure given by Kirchner [1999]; Lazar and Kirchner [2007a,b]. Let us take an arbitrary infinitesimal functional derivative $\delta U$ of the *elastic energy*

$$U = \int_V W \,\mathrm{d}V \,, \tag{23}$$

where $V$ is an arbitrary volume of the quasicrystalline material containing dislocations. Using Eq. (3), we obtain

$$\delta U = \frac{1}{2}\int_V \Big\{[\delta C_{ijkl}]\beta_{ij}^{\parallel}\beta_{kl}^{\parallel} + 2C_{ijkl}\beta_{ij}^{\parallel}[\delta\beta_{kl}^{\parallel}] + 2[\delta D_{ijBl}]\beta_{ij}^{\parallel}\beta_{Bl}^{\perp} + 2D_{ijBl}[\delta\beta_{ij}^{\parallel}]\beta_{Bl}^{\perp}$$
$$+ 2D_{ijBl}\beta_{ij}^{\parallel}[\delta\beta_{Bl}^{\perp}] + [\delta E_{AjBl}]\beta_{Aj}^{\perp}\beta_{Bl}^{\perp} + 2E_{AjBl}\beta_{Aj}^{\perp}[\delta\beta_{Bl}^{\perp}]\Big\}\,\mathrm{d}V \,. \tag{24}$$

Using the constitutive relations (7) and (8), Eq. (24) becomes

$$\delta U = \int_V \Big\{\sigma_{ij}^{\parallel}[\delta\beta_{ij}^{\parallel}] + \sigma_{Aj}^{\perp}[\delta\beta_{Aj}^{\perp}] + \frac{1}{2}[\delta C_{ijkl}]\beta_{ij}^{\parallel}\beta_{kl}^{\parallel} + [\delta D_{ijBl}]\beta_{ij}^{\parallel}\beta_{Bl}^{\perp}$$
$$+ \frac{1}{2}[\delta E_{AjBl}]\beta_{Aj}^{\perp}\beta_{Bl}^{\perp}\Big\}\,\mathrm{d}V \,. \tag{25}$$

### 3.1. *Eshelby stress tensor, configurational forces and J-integral*

For the derivation of a translational balance law, the functional derivative has to be translational

$$\delta = (\delta x_k)\partial_k \,, \tag{26}$$

where $(\delta x_k)$ is an infinitesimal shift in the $x_k$-direction and $\partial_k = \partial/\partial x_k$. The translational variation of the elastic energy (23) is given by

$$\delta U = \int_V \delta W \, \mathrm{d}V = \int_V [\partial_k W](\delta x_k) \, \mathrm{d}V = \int_V \partial_i [W\delta_{ik}](\delta x_k) \, \mathrm{d}V . \tag{27}$$

On the other hand, the translational variation of Eq. (25) reads

$$\delta U = \int_V \Big\{ \sigma_{ij}^\parallel [\partial_k \beta_{ij}^\parallel - \partial_j \beta_{ik}^\parallel] + \sigma_{ij}^\parallel \partial_j \beta_{ik}^\parallel + \sigma_{Aj}^\perp [\partial_k \beta_{Aj}^\perp - \partial_j \beta_{Ak}^\perp] + \sigma_{Aj}^\perp \partial_j \beta_{Ak}^\perp \tag{28}$$

$$+ \frac{1}{2} \beta_{ij}^\parallel [\partial_k C_{ijmn}] \beta_{mn}^\parallel + \beta_{ij}^\parallel [\partial_k D_{ijBn}] \beta_{Bn}^\perp + \frac{1}{2} \beta_{Aj}^\perp [\partial_k E_{AjBn}] \beta_{Bn}^\perp \Big\} (\delta x_k) \, \mathrm{d}V ,$$

where the second, third, fifth and sixth terms have been subtracted and added. The purpose now is to use the formulas for the dislocation density tensors

$$\epsilon_{kjl} \alpha_{il}^\parallel = \partial_k \beta_{ij}^\parallel - \partial_j \beta_{ik}^\parallel , \tag{29}$$

$$\epsilon_{kjl} \alpha_{Al}^\perp = \partial_k \beta_{Aj}^\perp - \partial_j \beta_{Ak}^\perp . \tag{30}$$

The third and sixth terms of Eq. (28) can be rewritten as follows

$$\sigma_{ij}^\parallel [\partial_j \beta_{ik}^\parallel] = \partial_j [\sigma_{ij}^\parallel \beta_{ik}^\parallel] - [\partial_j \sigma_{ij}^\parallel] \beta_{ik}^\parallel = \partial_j [\sigma_{ij}^\parallel \beta_{ik}^\parallel] + f_i^\parallel \beta_{ik}^\parallel , \tag{31}$$

$$\sigma_{Aj}^\perp [\partial_j \beta_{Ak}^\perp] = \partial_j [\sigma_{Aj}^\perp \beta_{Ak}^\perp] - [\partial_j \sigma_{Aj}^\perp] \beta_{Ak}^\perp = \partial_j [\sigma_{Aj}^\perp \beta_{Ak}^\perp] + f_A^\perp \beta_{Ak}^\perp , \tag{32}$$

where the equilibrium conditions (5) and (6) have been also used. Using Eq. (29)–(32), Eq. (28) becomes

$$\delta U = \int_V \Big\{ \partial_j [\sigma_{ij}^\parallel \beta_{ik}^\parallel + \sigma_{Aj}^\perp \beta_{Ak}^\perp] + \epsilon_{kjl} \sigma_{ij}^\parallel \alpha_{il}^\parallel + \epsilon_{kjl} \sigma_{Aj}^\perp \alpha_{Al}^\perp + f_i^\parallel \beta_{ik}^\parallel + f_A^\perp \beta_{Ak}^\perp \tag{33}$$

$$+ \frac{1}{2} \beta_{ij}^\parallel [\partial_k C_{ijmn}] \beta_{mn}^\parallel + \beta_{ij}^\parallel [\partial_k D_{ijBn}] \beta_{Bn}^\perp + \frac{1}{2} \beta_{Aj}^\perp [\partial_k E_{AjBn}] \beta_{Bn}^\perp \Big\} (\delta x_k) \mathrm{d}V .$$

Comparing Eqs. (27) and (33), we obtain the *global translational balance law for incompatible elasticity of quasicrystals*

$$\int_V \partial_j \big[ W\delta_{jk} - \sigma_{ij}^\parallel \beta_{ik}^\parallel - \sigma_{Aj}^\perp \beta_{Ak}^\perp \big] \mathrm{d}V = \int_V \Big\{ \epsilon_{kjl} \sigma_{ij}^\parallel \alpha_{il}^\parallel + \epsilon_{kjl} \sigma_{Aj}^\perp \alpha_{Al}^\perp + f_i^\parallel \beta_{ik}^\parallel + f_A^\perp \beta_{Ak}^\perp$$

$$+ \frac{1}{2} \beta_{ij}^\parallel [\partial_k C_{ijmn}] \beta_{mn}^\parallel + \beta_{ij}^\parallel [\partial_k D_{ijBn}] \beta_{Bn}^\perp + \frac{1}{2} \beta_{Aj}^\perp [\partial_k E_{AjBn}] \beta_{Bn}^\perp \Big\} \mathrm{d}V . \tag{34}$$

The integrand of the integral on the left-hand side of Eq. (34) is the divergence of the *Eshelby stress tensor for quasicrystals* (see also Agiasofitou *et al.* [2010]; Lazar and Agiasofitou [2014])

$$P_{kj} = W\delta_{jk} - \sigma_{ij}^\parallel \beta_{ik}^\parallel - \sigma_{Aj}^\perp \beta_{Ak}^\perp . \tag{35}$$

It should be emphasized that Eq. (35) is the Eshelby stress tensor valid for any quasicrystal. The integral on the right-hand side of Eq. (34) contains the terms breaking the translational symmetry and it is a sum of the so-called *configurational or material force densities*

$$f_k^{\text{conf}} = \epsilon_{kjl}\sigma_{ij}^{\|}\alpha_{il}^{\|} + \epsilon_{kjl}\sigma_{Aj}^{\perp}\alpha_{Al}^{\perp} + f_i^{\|}\beta_{ik}^{\|} + f_A^{\perp}\beta_{Ak}^{\perp}$$
$$+ \frac{1}{2}\,\beta_{ij}^{\|}[\partial_k C_{ijmn}]\beta_{mn}^{\|} + \beta_{ij}^{\|}[\partial_k D_{ijBn}]\beta_{Bn}^{\perp} + \frac{1}{2}\,\beta_{Aj}^{\perp}[\partial_k E_{AjBn}]\beta_{Bn}^{\perp}\,. \quad (36)$$

In particular, the first two terms in Eq. (36) give the *Peach-Koehler force density for quasicrystals* [Lazar and Agiasofitou, 2014]

$$f_k^{\text{PK}} = \epsilon_{kjl}\big(\sigma_{ij}^{\|}\alpha_{il}^{\|} + \sigma_{Aj}^{\perp}\alpha_{Al}^{\perp}\big)\,, \quad (37)$$

which is the material or configurational force density acting on the phonon dislocation density $\alpha_{il}^{\|}$ in the presence of a phonon stress $\sigma_{ij}^{\|}$ and on the phason dislocation density $\alpha_{Al}^{\perp}$ in the presence of a phason stress $\sigma_{Aj}^{\perp}$. The dynamical Peach-Koehler force density for quasicrystals has been derived by Agiasofitou *et al.* [2010]. The third and fourth terms in Eq. (36) give the configurational force density acting on a phonon body force density $f_i^{\|}$ in the presence of an elastic phonon distortion $\beta_{ik}^{\|}$ and on a phason body force density $f_A^{\perp}$ in the presence of an elastic phason distortion $\beta_{Ak}^{\perp}$

$$f_k^{\text{C}} = f_i^{\|}\beta_{ik}^{\|} + f_A^{\perp}\beta_{Ak}^{\perp}\,, \quad (38)$$

which is the *Cherepanov force density for quasicrystals* [Agiasofitou *et al.*, 2010]. Eq. (38) is the generalization of the classical Cherepanov force density (see, e.g., Cherepanov [1979, 1981]) known in elasticity theory towards quasicrystals. Finally, the last three terms in Eq. (36) constitute the *inhomogeneity force density* or *Eshelby force density for quasicrystals*

$$f_k^{\text{inh}} = \frac{1}{2}\,\beta_{ij}^{\|}[\partial_k C_{ijmn}]\beta_{mn}^{\|} + \beta_{ij}^{\|}[\partial_k D_{ijBn}]\beta_{Bn}^{\perp} + \frac{1}{2}\,\beta_{Aj}^{\perp}[\partial_k E_{AjBn}]\beta_{Bn}^{\perp}\,, \quad (39)$$

which appears due to the gradient of the constitutive tensors (see, e.g., Kirchner [1999]) when the quasicrystalline material is non-homogeneous. Using the definitions (37), (38), and (39), the configurational force density (36) can be written as

$$f_k^{\text{conf}} = f_k^{\text{PK}} + f_k^{\text{C}} + f_k^{\text{inh}}\,. \quad (40)$$

Using Eqs. (35) and (36) or (40), *the global translational balance law for incompatible elasticity of quasicrystals* can be written as

$$\int_V \partial_j P_{kj}\,\mathrm{d}V = \int_V f_k^{\text{conf}}\,\mathrm{d}V\,. \quad (41)$$

Finally, the divergence of the Eshelby stress tensor in Eq. (41) can be integrated out with the Gauss theorem yielding

$$J_k = \int_S P_{kj}\,\mathrm{d}S_j = \int_V f_k^{\text{conf}}\,\mathrm{d}V\,, \quad (42)$$

which represents the *vectorial **J**-integral for a quasicrystal* which is non-homogeneous in presence of dislocations and body forces. $S$ is the boundary surface of the volume $V$ and $\mathrm{d}S_j = n_j \mathrm{d}S$ with $\mathbf{n}$ being the outward unit normal vector to the surface $S$.

Furthermore, recalling the fact that a crack can always be represented as a dislocation distribution [Bilby and Eshelby, 1968; Lardner, 1974], the configurational force acting on a crack [Rice, 1968] in absence of external forces in an otherwise homogeneous material is nothing but the Peach-Koehler force on a dislocation distribution

$$J_k = \mathcal{F}_k^{\mathrm{PK}} = \int_V f_k^{\mathrm{PK}} \, \mathrm{d}V = \int_V \epsilon_{kjl} (\sigma_{ij}^{\parallel} \alpha_{il}^{\parallel} + \sigma_{Aj}^{\perp} \alpha_{Al}^{\perp}) \, \mathrm{d}V \, . \tag{43}$$

In other words, the Peach-Koehler force on a continuous distribution of dislocations generate the configurational force or the so-called driving force acting on a crack in a quasicrystal.

For a dislocation loop given by Eqs. (21) and (22) in absence of body forces in an otherwise homogeneous quasicrystal, the ***J**-integral* (42) reduces to the Peach-Koehler force

$$J_k = \mathcal{F}_k^{\mathrm{PK}} = \int_V f_k^{\mathrm{PK}} \, \mathrm{d}V = \oint_{\mathcal{C}} \epsilon_{kjm} (b_i^{\parallel} \sigma_{ij}^{\parallel} + b_A^{\perp} \sigma_{Aj}^{\perp}) \, \mathrm{d}l_m \, , \tag{44}$$

where the following relation [deWit, 1973] has been used

$$\int_V \delta_j(\mathcal{C}) \, f(\boldsymbol{x}) \, \mathrm{d}V = \oint_{\mathcal{C}} f(\boldsymbol{x}) \, \mathrm{d}l_j \, . \tag{45}$$

Eq. (44) gives the *Peach-Koehler force on a dislocation loop in a quasicrystal* with the phonon and phason components of the Burgers vector, $b_i^{\parallel}$ and $b_A^{\perp}$, under given phonon and phason stress fields, $\sigma_{ij}^{\parallel}$ and $\sigma_{Aj}^{\perp}$. Eq. (44) is in agreement with the Peach-Koehler force on a dislocation loop in a quasicrystal derived by Lazar and Agiasofitou [2014] using the hyperspace notation. The Peach-Koehler force between two dislocation loops in quasicrystals can be found in Lazar and Agiasofitou [2014].

For homogeneous, dislocation-free and source-free quasicrystals, the configurational force densities are vanished, $f_k^{\mathrm{inh}} = 0$, $f_k^{\mathrm{PK}} = 0$, and $f_k^C = 0$, respectively. Then, we recover the *Eshelby stress tensor for compatible elasticity of quasicrystals*

$$P_{kj} = W\delta_{jk} - \sigma_{ij}^{\parallel} u_{i,k}^{\parallel} - \sigma_{Aj}^{\perp} u_{A,k}^{\perp} \, , \tag{46}$$

which is divergence-less

$$P_{kj,j} = 0 \tag{47}$$

and thus $J_k = 0$. Eq. (46) is in agreement with the static version of the energy-momentum tensor given by Shi [2007].

### 3.2. *Scaling flux vector, configurational work and M-integral*

In this subsection, we specify the functional derivative to be dilatational

$$\delta = x_k \partial_k \,. \tag{48}$$

Taking the dilatational variation of the elastic energy (23) and following a similar procedure as in the Subsection 3.1, we get

$$\int_V x_k \partial_j P_{kj} \mathrm{d}V = \int_V x_k f_k^{\mathrm{conf}} \mathrm{d}V. \tag{49}$$

The first integral of Eq. (49) can be rewritten

$$\int_V x_k \partial_j P_{kj} \mathrm{d}V = \int_V \left[ \partial_j (x_k P_{kj}) - P_{kk} \right] \mathrm{d}V \,, \tag{50}$$

where the trace of the Eshelby stress tensor (35) is equal to

$$P_{kk} = \frac{d-2}{2} \left[ \beta_{ij}^{\parallel} \sigma_{ij}^{\parallel} + \beta_{Aj}^{\perp} \sigma_{Aj}^{\perp} \right], \tag{51}$$

and $d = \delta_{kk}$ is the space dimension. Thus, $d = 3$ for three dimensions and $d = 2$ for two dimensions. Using the additive decompositions (9) and the equilibrium Eqs. (5) and (6), Eq. (51) becomes

$$P_{kk} = \frac{d-2}{2} \left[ \partial_j \big( u_i^{\parallel} \sigma_{ij}^{\parallel} + u_A^{\parallel} \sigma_{Aj}^{\parallel} \big) + f_i^{\parallel} u_i^{\parallel} + f_A^{\perp} u_A^{\perp} - \sigma_{ij}^{\parallel} \beta_{ij}^{\parallel \mathrm{P}} - \sigma_{Aj}^{\perp} \beta_{Aj}^{\perp \mathrm{P}} \right]. \tag{52}$$

Finally, using Eqs. (50) and (52), Eq. (49) gives the *global balance law for scaling transformations for incompatible elasticity of quasicrystals*

$$\int_V \partial_j \left[ x_k P_{kj} - \frac{d-2}{2} \big( u_i^{\parallel} \sigma_{ij}^{\parallel} + u_A^{\perp} \sigma_{Aj}^{\perp} \big) \right] \mathrm{d}V$$
$$= \int_V \left\{ x_k f_k^{\mathrm{conf}} + \frac{d-2}{2} \left( f_i^{\parallel} u_i^{\parallel} + f_A^{\perp} u_A^{\perp} - \sigma_{ij}^{\parallel} \beta_{ij}^{\parallel \mathrm{P}} - \sigma_{Aj}^{\perp} \beta_{Aj}^{\perp \mathrm{P}} \right) \right\} \mathrm{d}V \,. \tag{53}$$

The integrand of the first integral in Eq. (53) is the divergence of the *dilatation or scaling flux vector for quasicrystals*

$$Y_j = x_k P_{kj} - \frac{d-2}{2} \big( u_i^{\parallel} \sigma_{ij}^{\parallel} + u_A^{\perp} \sigma_{Aj}^{\perp} \big) \,. \tag{54}$$

Eq. (54) represents the dilatation or scaling flux vector valid for any quasicrystal. In addition, it can be seen that the terms appearing in the integrand of the second integral in Eq. (53) break the scaling invariance and form the *total configurational work density*

$$w^{\mathrm{tot}} = w^{\mathrm{conf}} + w^{\mathrm{intr}}, \tag{55}$$

where

$$w^{\mathrm{conf}} = x_k f_k^{\mathrm{conf}} \tag{56}$$

is the *configurational work density* produced by the configurational force density (40) and

$$w^{\mathrm{intr}} = \frac{d-2}{2}\left(f_i^{\parallel}u_i^{\parallel} + f_A^{\perp}u_A^{\perp} - \sigma_{ij}^{\parallel}\beta_{ij}^{\parallel\mathrm{P}} - \sigma_{Aj}^{\perp}\beta_{Aj}^{\perp\mathrm{P}}\right) \tag{57}$$

is the *intrinsic* or *field work density* which is the sum of the work due to the phonon body force density vector, phason body force density vector, phonon plastic distortion tensor and phason plastic distortion tensor. The factor

$$d_{u^{\parallel}} = d_{u^{\perp}} = -\frac{d-2}{2} \tag{58}$$

is the *scaling* or *canonical dimension* for the phonon displacement vector $u_i^{\parallel}$ and the phason displacement vector $u_A^{\perp}$.

Using the definitions (54) and (55), the *global balance law for scaling transformations for quasicrystals* (53) can be written as

$$\int_V \partial_j Y_j \, \mathrm{d}V = \int_V w^{\mathrm{tot}} \, \mathrm{d}V \,. \tag{59}$$

The first integral of Eq. (59) can be transformed into a surface integral

$$M = \int_S Y_j \, \mathrm{d}S_j = \int_V w^{\mathrm{tot}} \, \mathrm{d}V \,, \tag{60}$$

which represents the *M-integral for incompatible elasticity of a quasicrystalline material* which is non-homogeneous, in presence of phonon and phason body forces. It becomes evident that the *M*-integral (60) has the physical meaning of the total configurational work including the configurational work produced by the considered configurational forces plus the intrinsic or field work produced by the body force density vectors and the plastic distortion tensors. For a relevant discussion about the physical meaning of the *M*-integral of dislocations in incompatible elasticity and its specific expression and physical meaning for straight dislocations, the reader is addressed to Agiasofitou and Lazar [2016].

From Eq. (53), for a homogeneous ($f_k^{\mathrm{inh}} = 0$), dislocation-free ($f_k^{\mathrm{PK}} = 0$; $\beta_{ij}^{\parallel\mathrm{P}} = 0$, $\beta_{Aj}^{\perp\mathrm{P}} = 0$) and source-free ($f_k^{\mathrm{C}} = 0$; $f_i^{\parallel} = 0$, $f_A^{\perp} = 0$) quasicrystalline material, we obtain a divergence-less scaling flux vector. Therefore, we have a conservation law, $Y_{j,j} = 0$, and a conserved *M*-integral, $M = 0$.

The space dimension influences greatly the precise expression of the *M*-integral. For three-dimensional dislocation problems ($d = 3$) of incompatible elasticity of non-homogeneous quasicrystals in presence of body forces, the *M*-integral (60) becomes

$$\begin{aligned}
M &= \int_S \left[x_k P_{kj} - \frac{1}{2}\left(u_i^{\parallel}\sigma_{ij}^{\parallel} + u_A^{\perp}\sigma_{Aj}^{\perp}\right)\right] \mathrm{d}S_j \\
&= \int_V \left\{x_k f_k^{\mathrm{conf}} + \frac{1}{2}\left(f_i^{\parallel}u_i^{\parallel} + f_A^{\perp}u_A^{\perp} - \sigma_{ij}^{\parallel}\beta_{ij}^{\parallel\mathrm{P}} - \sigma_{Aj}^{\perp}\beta_{Aj}^{\perp\mathrm{P}}\right)\right\} \mathrm{d}V \,. \tag{61}
\end{aligned}$$

One can recognize that the last two terms of the volume integral in Eq. (61) is the *dislocation energy* $U_{\mathrm{d}}$ for quasicrystals, which can be defined similar to the corresponding expression in incompatible elasticity (see, e.g., Mura [1987]; Agiasofitou and Lazar [2016]) as follows

$$U_{\mathrm{d}} = -\frac{1}{2} \int_V \left( \sigma_{ij}^{\parallel} \beta_{ij}^{\parallel \mathrm{P}} + \sigma_{Aj}^{\perp} \beta_{Aj}^{\perp \mathrm{P}} \right) \mathrm{d}V \,. \tag{62}$$

Eq. (62) is the dislocation energy in a quasicrystal caused by the phonon and phason plastic distortions, $\beta_{ij}^{\parallel \mathrm{P}}$ and $\beta_{Aj}^{\perp \mathrm{P}}$, of a dislocation in the presence of the phonon and phason stress fields, $\sigma_{ij}^{\parallel}$ and $\sigma_{Aj}^{\perp}$. Thus, it can be seen that the dislocation energy is one part of the $M$-integral. Moreover, the *interaction energy* between two dislocations in quasicrystals reads in terms of the dislocation energy (62) (see, e.g., [Mura, 1987; Agiasofitou and Lazar, 2016])

$$U_{\mathrm{int}} = 2U_{\mathrm{d}} \,. \tag{63}$$

The $M$-integral for dislocations (61) in an otherwise homogeneous material in absence of body forces reduces to

$$M = \int_S \left[ x_k P_{kj} - \frac{1}{2} \left( u_i^{\parallel} \sigma_{ij}^{\parallel} + u_A^{\perp} \sigma_{Aj}^{\perp} \right) \right] \mathrm{d}S_j = \int_V x_k f_k^{\mathrm{PK}} \, \mathrm{d}V + U_{\mathrm{d}} = \mathcal{W}^{\mathrm{PK}} + U_{\mathrm{d}} \,, \tag{64}$$

where $\mathcal{W}^{\mathrm{PK}}$ is the *configurational work produced by the Peach-Koehler force density*.

For a dislocation loop given by Eqs. (19)–(22) interacting with the stress fields, $\sigma_{ij}^{\parallel}$ and $\sigma_{Aj}^{\perp}$, in an otherwise homogeneous quasicrystalline material without body forces, the $M$-integral (64) takes the precise expression

$$M = \oint_{\mathcal{C}} \epsilon_{kjm} x_k \left( b_i^{\parallel} \sigma_{ij}^{\parallel} + b_A^{\perp} \sigma_{Aj}^{\perp} \right) \mathrm{d}l_m + \frac{1}{2} \int_{\mathcal{S}} \left( b_i^{\parallel} \sigma_{ij}^{\parallel} + b_A^{\perp} \sigma_{Aj}^{\perp} \right) \mathrm{d}S_j \,, \tag{65}$$

where above the following relation [deWit, 1973]

$$\int_V \delta_j(\mathcal{S}) \, f(\boldsymbol{x}) \, \mathrm{d}V = \int_{\mathcal{S}} f(\boldsymbol{x}) \, \mathrm{d}S_j \tag{66}$$

has been used. Eq. (65) is the total configurational work done by a dislocation loop and it consists of two parts, namely an integral over the dislocation line $\mathcal{C}$ and an integral over the dislocation surface $\mathcal{S}$. The first term in Eq. (65) is the configurational or material work done by the Peach-Koehler force density and the second term in Eq. (65) is the dislocation energy caused by a dislocation loop with phonon and phason components of the Burgers vector, $b_i^{\parallel}$ and $b_A^{\perp}$, in presence of given phonon and phason stress fields, $\sigma_{ij}^{\parallel}$ and $\sigma_{Aj}^{\perp}$,

$$U_{\mathrm{d}} = \frac{1}{2} \int_{\mathcal{S}} \left( b_i^{\parallel} \sigma_{ij}^{\parallel} + b_A^{\perp} \sigma_{Aj}^{\perp} \right) \mathrm{d}S_j \,. \tag{67}$$

It is obvious that for the calculation of the $M$-integral (65) only the phonon and phason stress fields, $\sigma_{ij}^{\parallel}$ and $\sigma_{Aj}^{\perp}$, are needed. These stress fields may be caused by another defect or by an external source (e.g., external force).

For two-dimensional problems ($d = 2$), the $M$-integral (60) of a quasicrystalline material which is non-homogeneous in presence of body forces simplifies to

$$M = \int_S x_k P_{kj}\, \mathrm{d}S_j = \int_V x_k f_k^{\mathrm{conf}}\, \mathrm{d}V \,, \qquad (68)$$

where $k, j = 1, 2$. In this case, the $M$-integral consists only of the configurational work produced by the configurational force density (40).

### 3.3. *Angular momentum tensor, configurational vector moments and L-integral*

In general, for quasicrystals, there are many kinds of symmetries of the material coefficients (see, e.g., [Ding *et al.*, 1993]). Since the rotational conservation laws in the material space depend directly on the constitutive equations with symmetries described by the material coefficients, we give here only some representative examples for rotational balance and conservation laws. In particular, we examine one-dimensional quasicrystals, three-dimensional cubic quasicrystals and two-dimensional decagonal quasicrystals. The fact that quasicrystals have crystallographic or non-crystallographic symmetries plays also an important role to the rotational conservation laws.

#### 3.3.1. *One-dimensional quasicrystals*

We consider one-dimensional quasicrystals which are periodic in the $xy$-plane and quasiperiodic in the $z$-direction. One-dimensional quasicrystals are quasicrystals with crystallographic symmetries. For one-dimensional quasicrystals, the phason displacement field is a scalar field, $u_A^\perp \equiv u_3^\perp$.

We specify here the functional derivative to be rotational in three dimensions, that means $SO(3)$-transformations

$$\delta = (\delta x_k)\epsilon_{kji} x_j \partial_i \,, \qquad (69)$$

where $(\delta x_k)$ denotes the $x_k$-direction of the axis of rotation. Using similar manipulations as in Subsection 3.1, we find

$$\int_V \epsilon_{kji} x_j [\partial_l P_{il}]\mathrm{d}V = \int_V \epsilon_{kji} x_j f_i^{\mathrm{conf}}\, \mathrm{d}V \,. \qquad (70)$$

The first integral of Eq. (70) is rewritten as follows

$$\int_V \epsilon_{kji} x_j [\partial_l P_{il}]\mathrm{d}V = \int_V \epsilon_{kji} [\partial_l (x_j P_{il}) - P_{ij}]\mathrm{d}V \,. \qquad (71)$$

It can be seen that in Eq. (71) the skew-symmetric part of the Eshelby stress tensor (35) is appearing, which for one-dimensional quasicrystals is

$$\epsilon_{kji} P_{ij} = -\epsilon_{kji}\big(\sigma_{lj}^{\parallel}\beta_{li}^{\parallel} + \sigma_{3j}^{\perp}\beta_{3i}^{\perp}\big) \,. \qquad (72)$$

In Eq. (70), we firstly use Eqs. (71) and (72), after we add and subtract the term $\epsilon_{kji}\sigma_{il}^{\|}\beta_{jl}^{\|}$, we use the additive decomposition for the phonon fields (9), and finally the equilibrium Eq. (5). Then, we obtain the *global rotational balance law for incompatible elasticity of one-dimensional quasicrystals*

$$\int_V \epsilon_{kji}\partial_l \big[x_j P_{il} + u_j^{\|}\sigma_{il}^{\|}\big]\mathrm{d}V$$

$$= \int_V \epsilon_{kji}\Big\{x_j f_i^{\mathrm{conf}} - u_j^{\|} f_i^{\|} + \beta_{jl}^{\|\mathrm{P}}\sigma_{il}^{\|} + \beta_{jl}^{\|}\sigma_{il}^{\|} + \beta_{lj}^{\|}\sigma_{li}^{\|} + \beta_{3j}^{\perp}\sigma_{3i}^{\perp}\Big\}\mathrm{d}V. \qquad (73)$$

The integrand of the first integral in Eq. (73) is the divergence of the *angular momentum tensor for one-dimensional quasicrystals*

$$M_{kl} = \epsilon_{kji}\big[x_j P_{il} + u_j^{\|}\sigma_{il}^{\|}\big] \qquad (74)$$

consisting of two parts

$$M_{kl} = M_{kl}^{(\mathrm{o})} + M_{kl}^{(\mathrm{i})}. \qquad (75)$$

The first one

$$M_{kl}^{(\mathrm{o})} = \epsilon_{kji} x_j P_{il} \qquad (76)$$

is the *orbital angular momentum tensor* given in terms of the Eshelby stress tensor (35) and the second one

$$M_{kl}^{(\mathrm{i})} = \epsilon_{kji} u_j^{\|}\sigma_{il}^{\|} \qquad (77)$$

is the *intrinsic* or *spin angular momentum tensor* given in terms only of the phonon displacement vector.

**Remark 3.1** We would like to remark that for one-dimensional quasicrystals, the phason displacement vector does not give a contribution to the spin angular momentum tensor. The reason is that the phason displacement field $u_A^{\perp} \equiv u_3^{\perp}$ transforms as a scalar field under spatial rotations. Due to this fact, there is also an "asymmetry" between the phonon and phason fields in the global rotational balance law (73). On the other hand, it is clear that even if the phason displacement field $u_3^{\perp}$ is not disturbed under the rotational transformations (or it is transformed as a scalar) the forms of the angular momentum tensor (74) and of the rotational balance law (73) are significantly affected by the phason components.

The integrand of the second integral in Eq. (73) is the *total configurational* or *material vector moment density*

$$m_k^{\mathrm{tot}} = m_k^{\mathrm{conf}} + m_k^{\mathrm{intr}} + m_k^{\mathrm{anis}} \qquad (78)$$

containing terms breaking the rotational symmetry. The first term

$$m_k^{\mathrm{conf}} = \epsilon_{kji} x_j f_i^{\mathrm{conf}} \qquad (79)$$

is the *configurational vector moment density* produced by the configurational force density $f_i^{\mathrm{conf}}$ given in Eq. (40). The following two terms constitute the *intrinsic* or *field vector moment density*

$$m_k^{\mathrm{intr}} = -\epsilon_{kji} u_j^{\|} f_i^{\|} + \epsilon_{kji} \beta_{jl}^{\|\mathrm{P}} \sigma_{il}^{\|} \tag{80}$$

due to the phonon body force density vector and the phonon plastic distortion tensor. The remaining terms form a *vector moment density due to the material anisotropy*

$$m_k^{\mathrm{anis}} = \epsilon_{kji} \big[ \beta_{jl}^{\|} \sigma_{il}^{\|} + \beta_{lj}^{\|} \sigma_{li}^{\|} + \beta_{3j}^{\perp} \sigma_{3i}^{\perp} \big] . \tag{81}$$

Finally, using the definitions (74) and (78), the *global rotational balance law for incompatible elasticity of one-dimensional quasicrystals* (73) reads

$$\int_V \partial_l M_{kl} \, \mathrm{d}V = \int_V m_k^{\mathrm{tot}} \, \mathrm{d}V . \tag{82}$$

Using the divergence theorem in Eq. (82), the volume integral can be rewritten as a surface integral

$$L_k = \int_S M_{kj} \, \mathrm{d}S_j = \int_V m_k^{\mathrm{tot}} \, \mathrm{d}V , \tag{83}$$

representing the *vectorial $\boldsymbol{L}$-integral of a one-dimensional quasicrystal*.

Using Eq. (66), the $\boldsymbol{L}$-integral (83) for a dislocation loop given by Eqs. (19)–(22) in an otherwise homogeneous one-dimensional quasicrystal in absence of body forces reduces to

$$L_k = \oint_{\mathcal{C}} x_j \big( b_i^{\|} \sigma_{ik}^{\|} \, \mathrm{d}l_j - b_i^{\|} \sigma_{ij}^{\|} \, \mathrm{d}l_k + b_3^{\perp} \sigma_{3k}^{\perp} \, \mathrm{d}l_j - b_3^{\perp} \sigma_{3j}^{\perp} \, \mathrm{d}l_k \big) - \int_{\mathcal{S}} \epsilon_{kji} b_j^{\|} \sigma_{il}^{\|} \, \mathrm{d}S_l$$
$$+ \int_V \epsilon_{kji} \big( \beta_{jl}^{\|} \sigma_{il}^{\|} + \beta_{lj}^{\|} \sigma_{li}^{\|} + \beta_{3j}^{\perp} \sigma_{3i}^{\perp} \big) \, \mathrm{d}V . \tag{84}$$

Eq. (84) represents the total configurational or material vector moment of a dislocation loop under given stress fields $\sigma_{ij}^{\|}$, $\sigma_{3j}^{\perp}$ and it consists of three parts. The first term (line integral) in Eq. (84) is the vector moment due to the Peach-Koehler force density, the second term (surface integral) represents the vector moment due to the phonon plastic distortion and the third term (volume integral) is the vector moment due to the material anisotropy.

We now return to discuss Eq. (81). Eq. (81) could lead to a so-called isotropy condition for one-dimensional quasicrystals. From the crystallographic point of view, looking at the 10 Laue classes of one-dimensional quasicrystals (see e.g., Wang *et al.* [1997]; Fan [2011]), and keeping in mind the Hermann theorem, it can be seen that only the two Laue classes, 9 and 10, of one-dimensional hexagonal quasicrystals can be isotropic, and this only in the basal plane. In what follows, we examine the isotropy of one-dimensional hexagonal quasicrystals of Laue class 10.

**Proposition 1** *For one-dimensional hexagonal quasicrystals of Laue class 10, the component of the vector moment density due to the material anisotropy about the*

$z$-axis, $m_3^{anis}$, equals zero:

$$m_3^{anis} = \epsilon_{3ji}\big[\beta_{jl}^{\parallel}\sigma_{il}^{\parallel} + \beta_{lj}^{\parallel}\sigma_{li}^{\parallel} + \beta_{3j}^{\perp}\sigma_{3i}^{\perp}\big] = 0\,, \quad i,j,l = 1,2,3. \tag{85}$$

*In other words, a one-dimensional hexagonal quasicrystal of Laue class 10 is isotropic in the basal xy-plane and Eq. (85) represents the corresponding isotropy condition.*

**Proof.** We substitute the explicit constitutive equations of one-dimensional hexagonal quasicrystals of Laue class 10 for the phonon and phason stresses into the isotropy condition (85). For one-dimensional hexagonal quasicrystals of Laue class 10, there are five independent phonon elastic constants $C_{11}$, $C_{12}$, $C_{13}$, $C_{33}$ and $C_{44}$ ($C_{66}$ is given by $2C_{66} = C_{11} - C_{12}$), two independent phason elastic constants $K_1$, $K_2$, and three independent phonon-phason coupling constants $R_1$, $R_2$ and $R_3$ [Wang *et al.*, 1997]. The non-vanishing components of the phonon stress tensor read (e.g., Wang *et al.* [1997]; Fan [2011])

$$\sigma_{11}^{\parallel} = C_{11}\beta_{11}^{\parallel} + C_{12}\beta_{22}^{\parallel} + C_{13}\beta_{33}^{\parallel} + R_1\beta_{33}^{\perp}\,, \tag{86}$$

$$\sigma_{22}^{\parallel} = C_{12}\beta_{11}^{\parallel} + C_{11}\beta_{22}^{\parallel} + C_{13}\beta_{33}^{\parallel} + R_1\beta_{33}^{\perp}\,, \tag{87}$$

$$\sigma_{33}^{\parallel} = C_{13}\beta_{11}^{\parallel} + C_{13}\beta_{22}^{\parallel} + C_{33}\beta_{33}^{\parallel} + R_2\beta_{33}^{\perp} \tag{88}$$

$$\sigma_{13}^{\parallel} = \sigma_{31}^{\parallel} = 2C_{44}\beta_{(13)}^{\parallel} + R_3\beta_{31}^{\perp}\,, \tag{89}$$

$$\sigma_{23}^{\parallel} = \sigma_{32}^{\parallel} = 2C_{44}\beta_{(23)}^{\parallel} + R_3\beta_{32}^{\perp}\,, \tag{90}$$

$$\sigma_{12}^{\parallel} = \sigma_{21}^{\parallel} = 2C_{66}\beta_{(12)}^{\parallel} \tag{91}$$

and the non-vanishing components of the phason stress tensor read (e.g., Wang *et al.* [1997]; Fan [2011])

$$\sigma_{33}^{\perp} = R_1(\beta_{11}^{\parallel} + \beta_{22}^{\parallel}) + R_2\beta_{33}^{\parallel} + K_1\beta_{33}^{\perp}\,, \tag{92}$$

$$\sigma_{31}^{\perp} = 2R_3\beta_{(31)}^{\parallel} + K_2\beta_{31}^{\perp}\,, \tag{93}$$

$$\sigma_{32}^{\perp} = 2R_3\beta_{(32)}^{\parallel} + K_2\beta_{32}^{\perp}\,, \tag{94}$$

where $\beta_{(ij)}^{\parallel} = 1/2(\beta_{ij}^{\parallel} + \beta_{ji}^{\parallel})$ is the elastic phonon strain tensor. If we substitute the corresponding stress components into the isotropy condition (85), then we obtain

$$\begin{aligned} m_3^{anis} &= \epsilon_{3ji}\big[\beta_{jl}^{\parallel}\sigma_{il}^{\parallel} + \beta_{lj}^{\parallel}\sigma_{li}^{\parallel} + \beta_{3j}^{\perp}\sigma_{3i}^{\perp}\big] \\ &= 2R_3\big(\beta_{(13)}^{\parallel}\beta_{32}^{\perp} - \beta_{(23)}^{\parallel}\beta_{31}^{\perp}\big) - 2R_3\big(\beta_{(31)}^{\parallel}\beta_{32}^{\perp} - \beta_{(32)}^{\parallel}\beta_{31}^{\perp}\big) = 0\,. \end{aligned} \tag{95}$$

Thus, the isotropy condition (85) for one-dimensional hexagonal quasicrystals of Laue class 10 is satisfied and a one-dimensional hexagonal quasicrystal of Laue class 10 is isotropic in the basal plane. □

**Remark 3.2** In the above Proof, it is interesting to observe that the phonon part of the isotropy condition (85) equals minus the phason part of the isotropy

condition

$$\epsilon_{3ji}\big[\beta_{jl}^{\parallel}\sigma_{il}^{\parallel} + \beta_{lj}^{\parallel}\sigma_{li}^{\parallel}\big] = -\epsilon_{3ji}\beta_{3j}^{\parallel}\sigma_{3i}^{\perp} = 2R_3\big(\beta_{(13)}^{\parallel}\beta_{32}^{\perp} - \beta_{(23)}^{\parallel}\beta_{31}^{\perp}\big)\,. \tag{96}$$

### 3.3.2. *Three-dimensional quasicrystals: cubic quasicrystals*

An important class of three-dimensional quasicrystals are the cubic quasicrystals. For three-dimensional cubic quasicrystals the phonon and phason displacement fields transform under the same vector representation [Yang *et al.*, 1993]. This has the advantage that we can interchange phonon and phason indices, and we can use small letters for both, the phonon and phason indices ($i = A = 1, 2, 3$). As we will see below, due to this property, the global rotational balance law, the $\boldsymbol{L}$-integral and the associated quantities will have a symmetric form in what concerns the phonon and phason fields.

Here, we investigate the $SO(3)$-transformations (69) for three-dimensional cubic quasicrystals. Eqs. (70) and (71) are also true for this case. Substituting Eq. (71) into Eq. (70), we have

$$\int_V \epsilon_{kji}\big[\partial_l(x_j P_{il}) - P_{ij}\big]\mathrm{d}V = \int_V \epsilon_{kji}x_j f_i^{\mathrm{conf}}\mathrm{d}V\,. \tag{97}$$

The skew-symmetric part of the Eshelby stress tensor for three-dimensional cubic quasicrystals reads

$$\epsilon_{kji}P_{ij} = -\epsilon_{kji}(\sigma_{lj}^{\parallel}\beta_{li}^{\parallel} + \sigma_{lj}^{\perp}\beta_{li}^{\perp})\,. \tag{98}$$

If we substitute Eq. (98) into Eq. (97), next we add and subtract the terms $\epsilon_{kji}\sigma_{il}^{\parallel}\beta_{jl}^{\parallel}$ and $\epsilon_{kji}\sigma_{il}^{\perp}\beta_{jl}^{\perp}$, we use the phonon and phason equilibrium equations (5) and (6), and the additive decompositions (9), then we obtain the *global rotational balance law for incompatible elasticity of three-dimensional cubic quasicrystals*

$$\int_V \epsilon_{kji}\partial_l\big[x_j P_{il} + u_j^{\parallel}\sigma_{il}^{\parallel} + u_j^{\perp}\sigma_{il}^{\perp}\big]\mathrm{d}V = \int_V \epsilon_{kji}\Big\{x_j f_i^{\mathrm{conf}} - u_j^{\parallel}f_i^{\parallel} - u_j^{\perp}f_i^{\perp}$$
$$+ \beta_{jl}^{\parallel\mathrm{P}}\sigma_{il}^{\parallel} + \beta_{jl}^{\perp\mathrm{P}}\sigma_{il}^{\perp} + \beta_{ji}^{\parallel}\sigma_{il}^{\parallel} + \beta_{lj}^{\parallel}\sigma_{li}^{\parallel} + \beta_{ji}^{\perp}\sigma_{il}^{\perp} + \beta_{lj}^{\perp}\sigma_{li}^{\perp}\Big\}\mathrm{d}V\,. \tag{99}$$

The *angular momentum tensor of three-dimensional cubic quasicrystals* reads

$$M_{kl} = \epsilon_{kji}\big[x_j P_{il} + u_j^{\parallel}\sigma_{il}^{\parallel} + u_j^{\perp}\sigma_{il}^{\perp}\big] \tag{100}$$

and consists of two parts

$$M_{kl} = M_{kl}^{(\mathrm{o})} + M_{kl}^{(\mathrm{i})}\,, \tag{101}$$

the *orbital angular momentum tensor*

$$M_{kl}^{(\mathrm{o})} = \epsilon_{kji}x_j P_{il} \tag{102}$$

and the *intrinsic* or *spin angular momentum tensor*

$$M_{kl}^{(\mathrm{i})} = \epsilon_{kji}(u_j^{\parallel}\sigma_{il}^{\parallel} + u_j^{\perp}\sigma_{il}^{\perp})\,, \tag{103}$$

which is given now in terms of both, the phonon and phason displacement vectors.

The integrand of the second integral in Eq. (99) is the *total configurational* or *material vector moment density*

$$m_k^{\mathrm{tot}} = m_k^{\mathrm{conf}} + m_k^{\mathrm{intr}} + m_k^{\mathrm{anis}} \,, \tag{104}$$

where

$$m_k^{\mathrm{conf}} = \epsilon_{kji} x_j f_i^{\mathrm{conf}} \,, \tag{105}$$

$$m_k^{\mathrm{intr}} = -\epsilon_{kji}(u_j^{\parallel} f_i^{\parallel} + u_j^{\perp} f_i^{\perp}) + \epsilon_{kji}(\beta_{jl}^{\parallel \mathrm{P}} \sigma_{il}^{\parallel} + \beta_{jl}^{\perp \mathrm{P}} \sigma_{il}^{\perp}) \,, \tag{106}$$

$$m_k^{\mathrm{anis}} = \epsilon_{kji}[\beta_{jl}^{\parallel} \sigma_{il}^{\parallel} + \beta_{lj}^{\parallel} \sigma_{li}^{\parallel} + \beta_{jl}^{\perp} \sigma_{il}^{\perp} + \beta_{lj}^{\perp} \sigma_{li}^{\perp}] \,. \tag{107}$$

The *intrinsic vector moment density*, Eq. (106), contains for cubic quasicrystals contributions from both, phonon and phason, body force density vectors and plastic distortion tensors. The *vector moment density due to the material anisotropy*, Eq. (107), is non-zero, since all the constitutive tensors are anisotropic for cubic quasicrystals. Namely, there are three elastic constants for the tensor $C_{ijkl}$, three elastic constants for the tensor $D_{ijkl}$, and three elastic constants for the tensor $E_{ijkl}$ [Yang *et al.*, 1993].

Using Eqs. (100) and (104), *the global rotational balance law for incompatible elasticity of three-dimensional cubic quasicrystals* (99) is written in the compact form

$$\int_V \partial_l M_{kl} \,\mathrm{d}V = \int_V m_k^{\mathrm{tot}} \,\mathrm{d}V \,, \tag{108}$$

giving rise to the *vectorial $\boldsymbol{L}$-integral for three-dimensional cubic quasicrystals*

$$L_k = \int_S M_{kj} \,\mathrm{d}S_j = \int_V m_k^{\mathrm{tot}} \,\mathrm{d}V \,. \tag{109}$$

For a three-dimensional cubic quasicrystal, the $\boldsymbol{L}$-integral (109) for a dislocation loop given by Eqs. (19)–(22) in an otherwise homogeneous material in absence of body forces reduces to

$$L_k = \oint_{\mathcal{C}} x_j (b_i^{\parallel} \sigma_{ik}^{\parallel} \,\mathrm{d}l_j - b_i^{\parallel} \sigma_{ij}^{\parallel} \,\mathrm{d}l_k + b_i^{\perp} \sigma_{ik}^{\perp} \,\mathrm{d}l_j - b_i^{\perp} \sigma_{ij}^{\perp} \,\mathrm{d}l_k) - \int_{\mathcal{S}} \epsilon_{kji}(b_j^{\parallel} \sigma_{il}^{\parallel} + b_j^{\perp} \sigma_{il}^{\perp}) \,\mathrm{d}S_l$$
$$+ \int_V \epsilon_{kji}(\beta_{jl}^{\parallel} \sigma_{il}^{\parallel} + \beta_{lj}^{\parallel} \sigma_{li}^{\parallel} + \beta_{jl}^{\perp} \sigma_{il}^{\perp} + \beta_{lj}^{\perp} \sigma_{li}^{\perp}) \,\mathrm{d}V \,. \tag{110}$$

### 3.3.3. *Two-dimensional quasicrystals: decagonal quasicrystals*

For two-dimensional quasicrystals, there are quasicrystals with crystallographic symmetries and quasicrystals with non-crystallographic symmetries. For crystallographic symmetries, both $\boldsymbol{u}^{\parallel}$ and $\boldsymbol{u}^{\perp}$ transform according to a vector-like representation. For quasicrystals with non-crystallographic symmetries, $\boldsymbol{u}^{\parallel}$ transforms according to a vector-like representation and $\boldsymbol{u}^{\perp}$ transforms according to another non-vector-like representation [Hu *et al.*, 1996]. Thus, for non-crystallographic symmetries the phason displacement can be interpreted as a tensor of higher rank since it does not transform under the vector representation. A decagonal quasicrystal is a quasicrystal with non-crystallographic symmetries [Zhou *et al.*, 2004].

We investigate the two-dimensional rotations for a decagonal quasicrystal with Laue class 14 and point groups $10mm$, $1022$, $\overline{10}m2$ and $10/mmm$ (see, e.g., Hu *et al.* [1996]). For two-dimensional quasicrystals, the phonon displacement vector has three components whereas the phason displacement vector has two components and both depend on the spatial coordinates in three dimensions. The coordinate system is chosen in such a way that the $x_3$-axis is along the periodic direction. For a two-dimensional decagonal quasicrystal, we use the fact that the rotation of the phason displacement field is three times the size of that of the phonon displacement field, which has been shown by Shi [2005].

We specify the functional derivative to be rotational in two dimensions, that means $SO(2)$-transformations around the $x_3$-axis

$$\delta = (\delta x_3)\epsilon_{3ji}x_j\partial_i \, , \tag{111}$$

where $(\delta x_3)$ denotes the $x_3$-direction of the axis of rotation. Following a similar procedure as in the previous Subsections 3.3.1 and 3.3.2, we obtain

$$\int_V \epsilon_{3ji}\big[\partial_t(x_j P_{il}) - P_{ij}\big]\mathrm{d}V = \int_V \epsilon_{3ji}x_j f_i^{\mathrm{conf}}\,\mathrm{d}V \, . \tag{112}$$

The skew-symmetric part of the Eshelby stress tensor (35) for two-dimensional decagonal quasicrystals reads

$$\epsilon_{3ji}P_{ij} = -\epsilon_{3ji}\big(\sigma_{lj}^{\parallel}\beta_{li}^{\parallel} + \sigma_{Aj}^{\perp}\beta_{Ai}^{\perp}\big)\, . \tag{113}$$

First, we substitute Eq. (113) into Eq. (112), after we add and subtract the terms $\epsilon_{3ji}\sigma_{il}^{\parallel}\beta_{jl}^{\parallel}$ and $3\epsilon_{3BA}\sigma_{Al}^{\perp}\beta_{Bl}^{\perp}$, we use the phonon and phason equilibrium equations (5) and (6), and the additive phonon and phason decompositions (9). The result of this derivation is the *global rotational balance law for incompatible elasticity of two-dimensional decagonal quasicrystals of Laue class 14*

$$\int_V \partial_t\big[\epsilon_{3ji}(x_j P_{il} + u_j^{\parallel}\sigma_{il}^{\parallel}) + 3\epsilon_{3BA}u_B^{\perp}\sigma_{Al}^{\perp}\big]\mathrm{d}V$$
$$= \int_V \Big\{\epsilon_{3ji}(x_j f_i^{\mathrm{conf}} - u_i^{\parallel}f_i^{\parallel} + \beta_{jl}^{\parallel\mathrm{P}}\sigma_{il}^{\parallel} + \beta_{jl}^{\parallel}\sigma_{il}^{\parallel} + \beta_{lj}^{\parallel}\sigma_{li}^{\parallel} + \beta_{Aj}^{\perp}\sigma_{Ai}^{\perp})$$
$$- 3\epsilon_{3BA}(u_B^{\perp}f_A^{\perp} - \beta_{Bl}^{\perp\mathrm{P}}\sigma_{Al}^{\perp} - \beta_{Bl}^{\perp}\sigma_{Al}^{\perp})\Big\}\mathrm{d}V\, , \qquad i,j,l = 1,2,3\, , \quad A,B = 1,2\, . \tag{114}$$

The integrand of the first integral in Eq. (114) is the divergence of the *angular momentum vector for a decagonal quasicrystal of Laue class 14*

$$M_l \equiv M_{3l} = \epsilon_{3ji}(x_j P_{il} + u_j^{\parallel}\sigma_{il}^{\parallel}) + 3\epsilon_{3BA}u_B^{\perp}\sigma_{Al}^{\perp}\, . \tag{115}$$

Eq. (115) consists of two parts

$$M_{3l} = M_{3l}^{(\mathrm{o})} + M_{3l}^{(\mathrm{i})}\, . \tag{116}$$

The first one

$$M_{3l}^{(\mathrm{o})} = \epsilon_{3ji} x_j P_{il} \tag{117}$$

is the *orbital angular momentum vector* given in terms of the Eshelby stress tensor (35) and the second one

$$M_{3l}^{(\mathrm{i})} = \epsilon_{3ji} u_j^{\|} \sigma_{il}^{\|} + 3\epsilon_{3BA} u_B^{\perp} \sigma_{Al}^{\perp} \tag{118}$$

is the *intrinsic* or *spin angular momentum vector* given in terms of the phonon displacement vector $u_j^{\|}$ and the phason displacement vector $u_B^{\perp}$ times 3.

The integrand of the second integral in Eq. (114) is the *total configurational* or *material moment density*

$$m_3^{\mathrm{tot}} = m_3^{\mathrm{conf}} + m_3^{\mathrm{intr}} + m_3^{\mathrm{anis}}, \tag{119}$$

where

$$m_3^{\mathrm{conf}} = \epsilon_{3ji} x_j f_i^{\mathrm{conf}}, \tag{120}$$

$$m_3^{\mathrm{intr}} = -\big(\epsilon_{3ji} u_j^{\|} f_i^{\|} + 3\epsilon_{3BA} u_B^{\perp} f_A^{\perp}\big) + \big(\epsilon_{3ji} \beta_{jl}^{\|\mathrm{P}} \sigma_{il}^{\|} + 3\epsilon_{3BA} \beta_{Bl}^{\perp\mathrm{P}} \sigma_{Al}^{\perp}\big), \tag{121}$$

$$m_3^{\mathrm{anis}} = \epsilon_{3ji}\big[\beta_{jl}^{\|}\sigma_{il}^{\|} + \beta_{lj}^{\|}\sigma_{li}^{\|} + \beta_{Aj}^{\perp}\sigma_{Ai}^{\perp}\big] + 3\epsilon_{3BA}\beta_{Bl}^{\perp}\sigma_{Al}^{\perp}. \tag{122}$$

**Proposition 2** *The configurational or material moment density due to the material anisotropy for two-dimensional decagonal quasicrystals of Laue class 14 is zero*

$$m_3^{anis} = \epsilon_{3ji}\big[\beta_{jl}^{\|}\sigma_{il}^{\|} + \beta_{lj}^{\|}\sigma_{li}^{\|} + \beta_{Aj}^{\perp}\sigma_{Ai}^{\perp}\big] + 3\epsilon_{3BA}\beta_{Bl}^{\perp}\sigma_{Al}^{\perp} = 0, \quad i,j,l = 1,2,3\,,$$
$$A,B = 1,2\,. \tag{123}$$

*In other words, a two-dimensional decagonal quasicrystal of Laue class 14 is isotropic in the xy-plane and Eq. (123) represents the corresponding isotropy condition.*

**Proof.** To prove that, we substitute the explicit constitutive equations of decagonal quasicrystals of Laue class 14 for the phonon and phason stresses into Eq. (123). For decagonal quasicrystals of Laue class 14, there are five independent phonon elastic constants $C_{11}$, $C_{12}$, $C_{13}$, $C_{33}$ and $C_{44}$ ($C_{66}$ is given by $2C_{66} = C_{11} - C_{22}$), three independent phason elastic constants $K_1$, $K_2$ and $K_3$, and one independent phonon-phason coupling constant $R$ [Lei *et al.*, 1999; Hu *et al.*, 2000; Edagawa and Takeuchi, 2007]. The non-vanishing components of the phonon stress tensor read

(e.g., Edagawa and Takeuchi [2007]; Shi [2005])

$$\sigma_{11}^{\parallel} = C_{11}\beta_{11}^{\parallel} + C_{12}\beta_{22}^{\parallel} + C_{13}\beta_{33}^{\parallel} + R\beta_{11}^{\perp} + R\beta_{22}^{\perp}, \tag{124}$$

$$\sigma_{22}^{\parallel} = C_{12}\beta_{11}^{\parallel} + C_{11}\beta_{22}^{\parallel} + C_{13}\beta_{33}^{\parallel} - R\beta_{11}^{\perp} - R\beta_{22}^{\perp}, \tag{125}$$

$$\sigma_{33}^{\parallel} = C_{13}\beta_{11}^{\parallel} + C_{13}\beta_{22}^{\parallel} + C_{33}\beta_{33}^{\parallel}, \tag{126}$$

$$\sigma_{12}^{\parallel} = \sigma_{21}^{\parallel} = 2C_{66}\beta_{(12)}^{\parallel} - R\beta_{12}^{\perp} + R\beta_{21}^{\perp}, \tag{127}$$

$$\sigma_{13}^{\parallel} = \sigma_{31}^{\parallel} = 2C_{44}\beta_{(13)}^{\parallel}, \tag{128}$$

$$\sigma_{23}^{\parallel} = \sigma_{32}^{\parallel} = 2C_{44}\beta_{(23)}^{\parallel} \tag{129}$$

and the non-vanishing components of the phason stress tensor are (e.g., Edagawa and Takeuchi [2007]; Shi [2005])

$$\sigma_{11}^{\perp} = K_1\beta_{11}^{\perp} + K_2\beta_{22}^{\perp} + R\beta_{11}^{\parallel} - R\beta_{22}^{\parallel}, \tag{130}$$

$$\sigma_{22}^{\perp} = K_2\beta_{11}^{\perp} + K_1\beta_{22}^{\perp} + R\beta_{11}^{\parallel} - R\beta_{22}^{\parallel}, \tag{131}$$

$$\sigma_{12}^{\perp} = K_1\beta_{12}^{\perp} - K_2\beta_{21}^{\perp} - 2R\beta_{(12)}^{\parallel}, \tag{132}$$

$$\sigma_{21}^{\perp} = -K_2\beta_{12}^{\perp} + K_1\beta_{21}^{\perp} + 2R\beta_{(12)}^{\parallel}, \tag{133}$$

$$\sigma_{13}^{\perp} = K_3\beta_{13}^{\perp}, \tag{134}$$

$$\sigma_{23}^{\perp} = K_3\beta_{23}^{\perp}. \tag{135}$$

By substituting the corresponding stress components into the isotropy condition (123), we obtain

$$\begin{aligned}
m_3^{\text{anis}} &= \epsilon_{3ji}\big[\beta_{jl}^{\parallel}\sigma_{il}^{\parallel} + \beta_{lj}^{\parallel}\sigma_{li}^{\parallel} + \beta_{Aj}^{\perp}\sigma_{Ai}^{\perp}\big] + 3\epsilon_{3BA}\beta_{Bl}^{\perp}\sigma_{Al}^{\perp} \\
&= -2R\big(\beta_{11}^{\parallel} - \beta_{22}^{\parallel}\big)\big(\beta_{12}^{\perp} - \beta_{21}^{\perp}\big) - 4R\beta_{(12)}^{\parallel}\big(\beta_{11}^{\perp} + \beta_{22}^{\perp}\big) \\
&\quad + 2R\big(\beta_{11}^{\parallel} - \beta_{22}^{\parallel}\big)\big(\beta_{12}^{\perp} - \beta_{21}^{\perp}\big) + 4R\beta_{(12)}^{\parallel}\big(\beta_{11}^{\perp} + \beta_{22}^{\perp}\big) = 0.
\end{aligned} \tag{136}$$

Thus, the isotropy condition (123) for decagonal quasicrystals of Laue class 14 is satisfied and therefore a decagonal quasicrystal of Laue class 14 is isotropic in the $xy$-plane. □

**Remark 3.3** In the above Proof, it is interesting to make the same observation as for the one-dimensional hexagonal quasicrystals; that the phonon part of the isotropy condition (123) equals minus the phason part of the isotropy condition

$$\begin{aligned}
\epsilon_{3ji}\big(\beta_{jl}^{\parallel}\sigma_{il}^{\parallel} + \beta_{lj}^{\parallel}\sigma_{li}^{\parallel}\big) &= -\big(\epsilon_{3ji}\beta_{Aj}^{\perp}\sigma_{Ai}^{\perp} + 3\epsilon_{3BA}\beta_{Bl}^{\perp}\sigma_{Al}^{\perp}\big) \\
&= -2R\big(\beta_{11}^{\parallel} - \beta_{22}^{\parallel}\big)\big(\beta_{12}^{\perp} - \beta_{21}^{\perp}\big) - 4R\beta_{(12)}^{\parallel}\big(\beta_{11}^{\perp} + \beta_{22}^{\perp}\big).
\end{aligned} \tag{137}$$

Therefore, the *total configurational* or *material moment density* (119) reduces to two non-vanishing terms

$$m_3^{\text{tot}} = m_3^{\text{conf}} + m_3^{\text{intr}}. \tag{138}$$

24 *M. Lazar & E. Agiasofitou*

Taking into account the Proposition 2, the final form of the *global rotational balance law for incompatible elasticity of two-dimensional decagonal quasicrystals of Laue class 14*, Eq. (114), becomes

$$\int_V \partial_l \big[\epsilon_{3ji}(x_j P_{il} + u_j^\parallel \sigma_{il}^\parallel) + 3\epsilon_{3BA} u_B^\perp \sigma_{Al}^\perp\big] \mathrm{d}V$$
$$= \int_V \Big\{\epsilon_{3ji}(x_j f_i^{\mathrm{conf}} - u_i^\parallel f_i^\parallel + \beta_{jl}^{\parallel\mathrm{P}} \sigma_{il}^\parallel) - 3\epsilon_{3BA}(u_B^\perp f_A^\perp - \beta_{Bl}^{\perp\mathrm{P}} \sigma_{Al}^\perp)\Big\} \mathrm{d}V\,,$$
$$i,j,l = 1,2,3\,, \quad A,B = 1,2\,, \tag{139}$$

or using Eqs. (115) and (138),

$$\int_V \partial_j M_{3j}\, \mathrm{d}V = \int_V m_3^{\mathrm{tot}}\, \mathrm{d}V\,. \tag{140}$$

The *$L_3$-integral for two-dimensional decagonal quasicrystals of Laue class 14* is given by

$$L_3 = \int_S M_{3j}\, \mathrm{d}S_j = \int_V m_3^{\mathrm{tot}}\, \mathrm{d}V\,. \tag{141}$$

For two-dimensional decagonal quasicrystals of Laue class 14, the $L_3$-integral (141) for a dislocation loop given by Eqs. (19)–(22) in an otherwise homogeneous material in absence of body forces reduces to

$$L_3 = \oint_{\mathcal{C}} x_j \big(b_i^\parallel \sigma_{i3}^\parallel\, \mathrm{d}l_j - b_i^\parallel \sigma_{ij}^\parallel\, \mathrm{d}l_3 + b_A^\perp \sigma_{A3}^\perp\, \mathrm{d}l_j - b_A^\perp \sigma_{Aj}^\perp\, \mathrm{d}l_3\big)$$
$$- \int_{\mathcal{S}} \big(\epsilon_{3ji} b_j^\parallel \sigma_{il}^\parallel + 3\epsilon_{3BA} b_B^\perp \sigma_{Al}^\perp\big)\, \mathrm{d}S_l\,. \tag{142}$$

## 4. *J*-, *M*-, and *L*-integrals of straight dislocations in quasicrystals

In this section, we determine the *J*-, *M*-, and *L*-integrals of a straight dislocation at the position $(\bar{x}, \bar{y})$ in the stress field of another straight dislocation at the origin of the coordinate system $(0,0)$ in absence of body forces in an otherwise homogeneous material. The *J*- and *M*-integrals are derived for arbitrary quasicrystals, whereas the *L*-integral due to its special characteristic features is given only for one-dimensional quasicrystals as a representative example.

For a straight dislocation, the dislocation line $\mathcal{C}$ is a straight line and the dislocation surface $\mathcal{S}$ is a semi-infinite plane bounded by $\mathcal{C}$. In particular, let $\mathcal{C}$ run along the $z$-axis and $\mathcal{S}$ be the part of the $xz$-plane for negative $x$ (see Fig. 2). Then, the only non-vanishing components of the phonon and phason plastic distortion tensors and phonon and phason dislocation density tensors for a straight dislocation at the

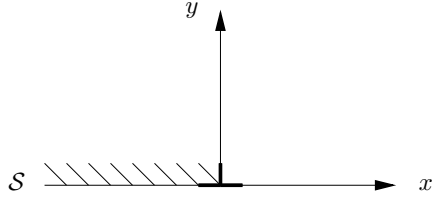

Figure 2.   Geometry of a straight dislocation at the position $(0,0)$

position $(\bar{x}, \bar{y})$ are

$$\beta_{i2}^{\parallel \mathrm{P}}(x-\bar{x}, y-\bar{y}) = b_i^{\parallel} H(-(x-\bar{x}))\,\delta(y-\bar{y})$$

$$= b_i^{\parallel}\,\delta(y-\bar{y})\int_{-\infty}^{0}\delta(x-\bar{x}-x')\,\mathrm{d}x'\,, \tag{143}$$

$$\beta_{A2}^{\perp \mathrm{P}}(x-\bar{x}, y-\bar{y}) = b_A^{\perp} H(-(x-\bar{x}))\,\delta(y-\bar{y})$$

$$= b_A^{\perp}\,\delta(y-\bar{y})\int_{-\infty}^{0}\delta(x-\bar{x}-x')\,\mathrm{d}x'\,, \tag{144}$$

$$\alpha_{i3}^{\parallel}(x-\bar{x}, y-\bar{y}) = b_i^{\parallel}\,\delta(x-\bar{x})\,\delta(y-\bar{y})\,, \tag{145}$$

$$\alpha_{A3}^{\perp}(x-\bar{x}, y-\bar{y}) = b_A^{\perp}\,\delta(x-\bar{x})\,\delta(y-\bar{y})\,, \tag{146}$$

where $\delta(.)$ and $H(.)$ denote the Dirac delta function and the Heaviside step function, respectively. The phonon and phason plastic distortions, $\beta_{i2}^{\parallel \mathrm{P}}$ and $\beta_{A2}^{\perp \mathrm{P}}$, possess a discontinuity on $\mathcal{S}$.

In addition, the dislocation energy (62) of straight dislocations in arbitrary quasicrystals takes the form

$$U_{\mathrm{d}} = \frac{1}{2}\,U_{\mathrm{int}} = -\frac{l_3}{2}\int_{-\infty}^{0}\Big(b_i^{\parallel}\sigma_{i2}^{\parallel}(\bar{x}+x', \bar{y}) + b_A^{\perp}\sigma_{A2}^{\perp}(\bar{x}+x', \bar{y})\Big)\,\mathrm{d}x'\,. \tag{147}$$

If we substitute Eqs. (143)–(146) into Eqs. (42), (64) and (83) and perform the volume integration, we obtain the ***J*-, *M*-, and *L*-integrals of straight dislocations**

$$J_k = \mathcal{F}_k^{\mathrm{PK}} = \epsilon_{kj3}\Big(b_i^{\parallel}\sigma_{ij}^{\parallel}(\bar{x}, \bar{y}) + b_A^{\perp}\sigma_{Aj}^{\perp}(\bar{x}, \bar{y})\Big)\,l_3\,, \tag{148}$$

$$M = \mathcal{W}^{\mathrm{PK}} + U_{\mathrm{d}} = \bar{x}_k J_k - \frac{l_3}{2}\int_{-\infty}^{0}\Big(b_i^{\parallel}\sigma_{i2}^{\parallel}(\bar{x}+x', \bar{y}) + b_A^{\perp}\sigma_{A2}^{\perp}(\bar{x}+x', \bar{y})\Big)\,\mathrm{d}x'\,, \tag{149}$$

$$L_k = \epsilon_{kji}\bar{x}_j J_i + l_3\int_{-\infty}^{0}\epsilon_{kji}b_j^{\parallel}\sigma_{i2}^{\parallel}(\bar{x}+x', \bar{y})\mathrm{d}x' + \int_V \epsilon_{kji}\big(\beta_{jl}^{\parallel}\sigma_{il}^{\parallel} + \beta_{lj}^{\parallel}\sigma_{li}^{\parallel} + \beta_{3j}^{\perp}\sigma_{3i}^{\perp}\big)\mathrm{d}V. \tag{150}$$

In the above equations, $l_3$ denotes the length of the dislocation line and it arises from the performance of $z$-integration. The ***J*-integral** given by Eq. (148) represents the Peach-Koehler force for straight dislocations in arbitrary quasicrystals in statics and it is in agreement with Li and Fan [1999]. The dynamical Peach-Koehler force for straight dislocations in arbitrary quasicrystals has been derived by Agiasofitou *et al.* [2010]. Eq. (149) represents the $M$-integral of straight dislocations in arbitrary quasicrystals. Eq. (150) is the $L$-integral of straight dislocations specified to one-dimensional quasicrystals. We can see in Eqs. (149) and (150) that the first parts of the $M$- and $L$-integrals are a direct contribution of the $J$-integral. The second terms in Eqs. (149) and (150) are due to the phonon and phason plastic distortions of a straight dislocation. The third term in Eq.(150) is due to the material anisotropy.

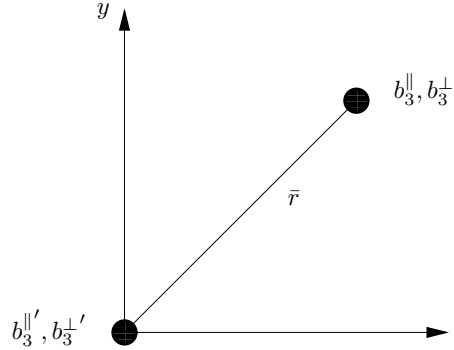

Figure 3.   Interaction between two parallel screw dislocations with Burgers vectors in $z$-direction.

**Remark 4.1**   Note that the position vector $\bar{x}_k$ in Eqs. (149) and (150) is obtained from the position of the phonon and phason dislocation density tensors (145) and (146) through the volume integration.

### 4.1. *Parallel screw dislocations in one-dimensional hexagonal quasicrystals*

In this subsection, we calculate the explicit form of the $J$-, $M$-, and $L$-integrals for two parallel screw dislocations with Burgers vectors $(b_3^{\parallel}, b_3^{\perp})$ and $(b_3^{\parallel\,\prime}, b_3^{\perp\,\prime})$ in a one-dimensional hexagonal quasicrystal (see Fig. 3) and we provide some fundamental relations between the $J$- and $M$-integrals.

The phonon and phason displacement vectors of a screw dislocation at the position $(0,0)$ possessing the Burgers vectors $(b_3^{\parallel\,\prime}, b_3^{\perp\,\prime})$ are [Li and Fan, 1999; Fan, 2011]

$$u_3^{\parallel}(x,y) = \frac{b_3^{\parallel\,\prime}}{2\pi} \arctan \frac{y}{x}\,, \qquad u_3^{\perp}(x,y) = \frac{b_3^{\perp\,\prime}}{2\pi} \arctan \frac{y}{x}\,. \qquad (151)$$

Note that the function $\arctan(y/x)$, as a single-valued function, possesses along the straight line $x = (-\infty, 0)$ a jump from $\pi$ to $-\pi$ when crossing the cut, as shown in Fig. 4. The cut corresponds to the dislocation surface $\mathcal{S}$. In the sense of generalized

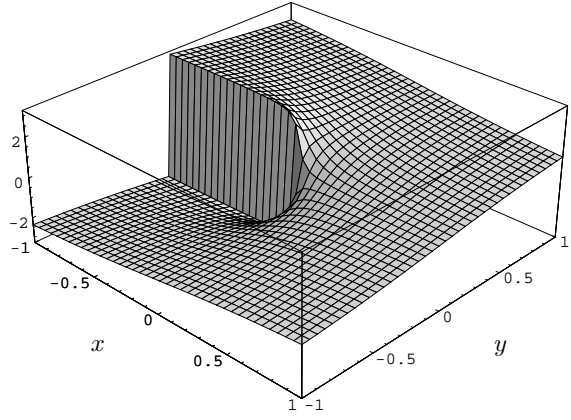

Figure 4.   The single-valued function $\arctan(y/x)$ with a jump.

functions [Gel'fand and Shilov, 1964], the gradient of the function $\arctan(y/x)$ reads

$$\left(\arctan\frac{y}{x}\right)_{,x} = -\frac{y}{x^2+y^2}\,, \tag{152}$$

$$\left(\arctan\frac{y}{x}\right)_{,y} = \frac{x}{x^2+y^2} + 2\pi H(-x)\delta(y)\,. \tag{153}$$

The discontinuity of the function $\arctan(y/x)$ leads to the second part in Eq. (153). Using Eqs. (152), (153) and (9), the displacement fields (151) give the phonon and phason plastic distortion tensors of a screw dislocation

$$\beta_{32}^{\|\mathrm{P}} = b_3^{\|'}H(-x)\delta(y)\,, \qquad \beta_{32}^{\perp\mathrm{P}} = b_3^{\perp'}H(-x)\delta(y)\,. \tag{154}$$

Eventually, the non-vanishing components of the phonon and phason elastic distortion tensors are

$$\beta_{31}^{\|}(x,y) = -\frac{b_3^{\|'}}{2\pi}\frac{y}{x^2+y^2}\,, \qquad \beta_{32}^{\|}(x,y) = \frac{b_3^{\|'}}{2\pi}\frac{x}{x^2+y^2}\,, \tag{155}$$

$$\beta_{31}^{\perp}(x,y) = -\frac{b_3^{\perp'}}{2\pi}\frac{y}{x^2+y^2}\,, \qquad \beta_{32}^{\perp}(x,y) = \frac{b_3^{\perp'}}{2\pi}\frac{x}{x^2+y^2}\,. \tag{156}$$

Substituting Eqs. (155) and (156) into Eqs. (89), (90), (93) and (94), the corresponding non-vanishing components of the phonon and phason stress tensors of a screw dislocation in a one-dimensional hexagonal quasicrystal are obtained (see

also [Li and Fan, 1999])

$$\sigma_{31}^{\parallel}(x,y) = -\frac{C_{44}b_3^{\parallel'} + Rb_3^{\perp'}}{2\pi}\frac{y}{x^2+y^2}\,, \qquad \sigma_{32}^{\parallel}(x,y) = \frac{C_{44}b_3^{\parallel'} + Rb_3^{\perp'}}{2\pi}\frac{x}{x^2+y^2}\,, \tag{157}$$

$$\sigma_{31}^{\perp}(x,y) = -\frac{Kb_3^{\perp'} + Rb_3^{\parallel'}}{2\pi}\frac{y}{x^2+y^2}\,, \qquad \sigma_{32}^{\perp}(x,y) = \frac{Kb_3^{\perp'} + Rb_3^{\parallel'}}{2\pi}\frac{x}{x^2+y^2}\,, \tag{158}$$

where $R = R_3$ and $K = K_2$.

### 4.1.1. *$\boldsymbol{J}$-integral*

Substituting Eqs. (157) and (158) into the formula of the $\boldsymbol{J}$-integral of straight dislocations, Eq. (148), we obtain the *$J_1$-, and $J_2$-integrals of two parallel screw dislocations in a one-dimensional hexagonal quasicrystal* per unit dislocation length $l_3$

$$\begin{aligned}\frac{J_1}{l_3} = \frac{\mathcal{F}_1^{\text{PK}}}{l_3} &= b_3^{\parallel}\sigma_{32}^{\parallel}(\bar{x},\bar{y}) + b_3^{\perp}\sigma_{32}^{\perp}(\bar{x},\bar{y}) \\ &= \frac{b_3^{\parallel}\big(C_{44}b_3^{\parallel'} + Rb_3^{\perp'}\big) + b_3^{\perp}\big(Kb_3^{\perp'} + Rb_3^{\parallel'}\big)}{2\pi}\frac{\bar{x}}{\bar{x}^2+\bar{y}^2}\,,\end{aligned} \tag{159}$$

$$\begin{aligned}\frac{J_2}{l_3} = \frac{\mathcal{F}_2^{\text{PK}}}{l_3} &= -b_3^{\parallel}\sigma_{31}^{\parallel}(\bar{x},\bar{y}) - b_3^{\perp}\sigma_{31}^{\perp}(\bar{x},\bar{y}) \\ &= \frac{b_3^{\parallel}\big(C_{44}b_3^{\parallel'} + Rb_3^{\perp'}\big) + b_3^{\perp}\big(Kb_3^{\perp'} + Rb_3^{\parallel'}\big)}{2\pi}\frac{\bar{y}}{\bar{x}^2+\bar{y}^2}\,.\end{aligned} \tag{160}$$

Thus, the $J_1$-, and $J_2$-integrals give the two non-vanishing components of the Peach-Koehler force between two screw dislocations and depend only on the position $(\bar{x},\bar{y})$ of the dislocation with Burgers vector $(b_3^{\parallel}, b_3^{\perp})$ in the stress field of the screw dislocation with Burgers vector $(b_3^{\parallel'}, b_3^{\perp'})$ at the position $(0,0)$.

In cylindrical coordinates, the two components $J_r$ and $J_\varphi$ of the $\boldsymbol{J}$-integral read

$$\frac{J_r}{l_3} = \frac{\mathcal{F}_r^{\text{PK}}}{l_3} = \frac{J_1\cos\varphi + J_2\sin\varphi}{l_3} = \frac{b_3^{\parallel}\big(C_{44}b_3^{\parallel'} + Rb_3^{\perp'}\big) + b_3^{\perp}\big(Kb_3^{\perp'} + Rb_3^{\parallel'}\big)}{2\pi}\frac{1}{\bar{r}}\,, \tag{161}$$

$$\frac{J_\varphi}{l_3} = \frac{\mathcal{F}_\varphi^{\text{PK}}}{l_3} = \frac{J_1\cos\varphi - J_2\sin\varphi}{l_3} = 0\,, \tag{162}$$

where $\bar{r} = \sqrt{\bar{x}^2+\bar{y}^2}$ is the distance between the two screw dislocations. It can be seen in Eqs. (161) and (162) that the Peach-Koehler force between two parallel screw dislocations in a one-dimensional hexagonal quasicrystalline material is purely radial as in the classical dislocation theory of elastic materials. The Peach-Koehler force is attractive or repulsive depending on the Burgers vectors as well as on the values of the material constants (see also relative discussion after Eq. (174)). Numerical values of the material constants $C_{44}$, $K$, and $R$ can be found in Li [2014] for a particular one-dimensional hexagonal quasicrystal. Moreover, the Peach-Koehler force contains phonon, phason as well as phonon-phason coupling contributions.

### 4.1.2. *M-integral*

For two parallel screw dislocations, the *M*-integral (149) reads

$$M = \mathcal{W}^{\mathrm{PK}} + U_{\mathrm{d}} = \bar{x}J_1 + \bar{y}J_2 - \frac{l_3}{2} \int_{-\infty}^{0} \left( b_3^{\parallel} \sigma_{32}^{\parallel}(\bar{x} + x', \bar{y}) + b_3^{\perp} \sigma_{32}^{\perp}(\bar{x} + x', \bar{y}) \right) \mathrm{d}x' . \tag{163}$$

Substituting the $J_1$-integral (159), the $J_2$-integral (160) as well as the phonon stress field $\sigma_{32}^{\parallel}$ and phason stress field $\sigma_{32}^{\perp}$ from Eqs. (157) and (158), respectively, into Eq. (163), the *M*-integral per unit dislocation length is given by

$$\frac{M}{l_3} = \frac{\mathcal{W}^{\mathrm{PK}}}{l_3} + \frac{U_{\mathrm{d}}}{l_3} \tag{164}$$

$$= \frac{b_3^{\parallel} \left( C_{44} b_3^{\parallel'} + R b_3^{\perp'} \right) + b_3^{\perp} \left( K b_3^{\perp'} + R b_3^{\parallel'} \right)}{2\pi} \left[ 1 - \frac{1}{2} \int_{-\infty}^{0} \frac{\bar{x} + x'}{(\bar{x} + x')^2 + \bar{y}^2} \, \mathrm{d}x' \right].$$

We carry out the integral part in Eq. (164)

$$\int_{-\infty}^{0} \frac{\bar{x} + x'}{(\bar{x} + x')^2 + \bar{y}^2} \, \mathrm{d}x' = \frac{1}{2} \ln((\bar{x} + x')^2 + \bar{y}^2) \Big|_{-\infty}^{0} = \ln \frac{\bar{r}}{L} , \tag{165}$$

where the limit to infinity is replaced by a finite number $L$, which corresponds to the size of the dislocated body. Using Eq. (165), Eq. (164) gives the final result for the *M-integral of two parallel screw dislocations* per unit dislocation length

$$\frac{M}{l_3} = \frac{\mathcal{W}^{\mathrm{PK}}}{l_3} + \frac{U_{\mathrm{d}}}{l_3} = K_s \left[ 2 - \ln \frac{\bar{r}}{L} \right] , \tag{166}$$

where

$$K_s = \frac{b_3^{\parallel} \left( C_{44} b_3^{\parallel'} + R b_3^{\perp'} \right) + b_3^{\perp} \left( K b_3^{\perp'} + R b_3^{\parallel'} \right)}{4\pi} \tag{167}$$

is the *pre-logarithmic energy factor for screw dislocations in a one-dimensional hexagonal quasicrystal*. As we can see, the pre-logarithmic energy factor contains phonon, phason as well as phonon-phason coupling contributions. For the pre-logarithmic energy factor for screw dislocations in classical elasticity theory of dislocations, the reader is addressed to [Hirth and Lothe, 1982; Teodosiu, 1982].

We now proceed to make a qualitative analysis of the *M*-integral (166). From Eq. (166), we obtain that the *configurational work done by the Peach-Koehler force between two screw dislocations* per unit dislocation length equals twice the pre-logarithmic energy factor

$$\frac{\mathcal{W}^{\mathrm{PK}}}{l_3} = \frac{\bar{x}_k J_k}{l_3} = 2K_s , \tag{168}$$

30   *M. Lazar & E. Agiasofitou*

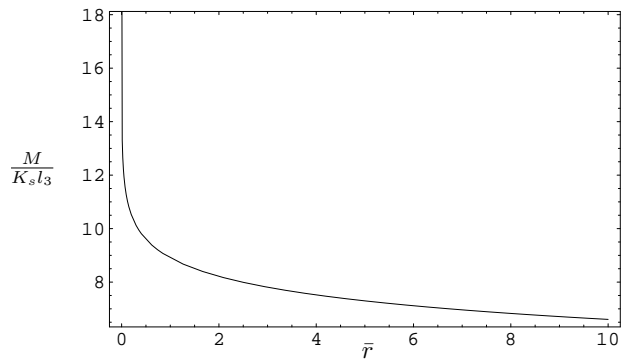

Figure 5.   Normalized $M$-integral of two screw dislocations in one-dimensional hexagonal quasicrystals against the distance $\bar{r}$ with $L = 1000$ for $K_s > 0$.

which is constant. In addition, *the dislocation energy of two parallel screw dislocations* per unit dislocation length is given by

$$\frac{U_{\mathrm{d}}}{l_3} = -K_s \ln \frac{\bar{r}}{L}\,, \tag{169}$$

which is a logarithmic term depending only on the distance $\bar{r}$ between the two screw dislocations for a given extension $L$ of the quasicrystal.

Using Eq. (63), the $M$-integral (166) can be rewritten in terms of the interaction energy as follows

$$\frac{M}{l_3} = 2K_s + \frac{1}{2}\frac{U_{\mathrm{int}}}{l_3}\,, \tag{170}$$

where

$$\frac{U_{\mathrm{int}}}{l_3} = -2K_s \ln \frac{\bar{r}}{L} \tag{171}$$

is the interaction energy between the two screw dislocations in a one-dimensional hexagonal quasicrystal, namely between the screw dislocation at $(0,0)$ with Burgers vector $(b_3^{\parallel'}, b_3^{\perp'})$ acting on the screw dislocation at $(\bar{x}, \bar{y})$ with Burgers vector $(b_3^{\parallel}, b_3^{\perp})$.

Eq. (170) reveals the physical meaning of the $M$-integral stating that the $M$-integral of two parallel screw dislocations (per unit dislocation length) is the half of the interaction energy between the two screw dislocations (per unit dislocation length) plus a constant term (namely twice the pre-logarithmic energy factor). The same result holds originally for dislocations in the framework of isotropic elasticity as it has been recently proved by Agiasofitou and Lazar [2016] and it is interesting to see that it is also true for dislocations in quasicrystals. The only difference lies in the precise expressions of the pre-logarithmic energy factor $K_s$ and of the interaction energy.

The normalized $M$-integral of two parallel screw dislocations is plotted in Fig. 5 and Fig. 6 for the cases $K_s > 0$ and $K_s < 0$, respectively. It can be seen that in the case $K_s > 0$, the $M$-integral decreases when $\bar{r}$ increases; and in the case $K_s < 0$, the $M$-integral is increasing when $\bar{r}$ increases.

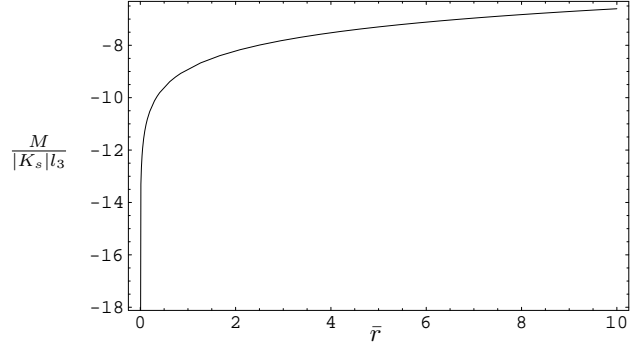

Figure 6. Normalized $M$-integral of two screw dislocations in one-dimensional hexagonal quasicrystals against the distance $\bar{r}$ with $L = 1000$ for $K_s < 0$.

Moreover, it is interesting to observe that the non-vanishing components of the $\boldsymbol{J}$-integral (see Eqs. (159), (160), (161)) can be simply expressed in terms of the pre-logarithmic energy factor $K_s$ as follows

$$\frac{J_1}{l_3} = \frac{\mathcal{F}_1^{\text{PK}}}{l_3} = 2K_s \frac{\bar{x}}{\bar{x}^2 + \bar{y}^2} \,, \tag{172}$$

$$\frac{J_2}{l_3} = \frac{\mathcal{F}_2^{\text{PK}}}{l_3} = 2K_s \frac{\bar{y}}{\bar{x}^2 + \bar{y}^2} \,, \tag{173}$$

$$\frac{J_r}{l_3} = \frac{\mathcal{F}_r^{\text{PK}}}{l_3} = 2K_s \frac{1}{\bar{r}} \,. \tag{174}$$

From Eq. (174), we see that the Peach-Koehler force is attractive if $K_s < 0$ and repulsive if $K_s > 0$.

### 4.1.3. $L_3$-integral

The $L_3$-integral for one-dimensional quasicrystals arising from Eq. (150) reads

$$L_3 = \epsilon_{3ji}\bar{x}_j J_i + l_3 \int_{-\infty}^{0} \epsilon_{3ji} b_j^{\parallel} \sigma_{i2}^{\parallel}(\bar{x} + x', \bar{y}) \, \mathrm{d}x' + \int_V \epsilon_{3ji}\big(\beta_{jl}^{\parallel}\sigma_{il}^{\parallel} + \beta_{lj}^{\parallel}\sigma_{li}^{\parallel} + \beta_{3j}^{\perp}\sigma_{3i}^{\perp}\big)\mathrm{d}V. \tag{175}$$

The second integral in Eq. (175) is zero, since we consider screw dislocations and the third integral is nothing but the integral of the $m_3^{\text{anis}}$ which is zero for one-dimensional hexagonal quasicrystals, which are isotropic in the basal plane as we have shown in the Proposition 1. Therefore,

$$L_3 = \epsilon_{3ji}\bar{x}_j J_i = \bar{x}J_2 - \bar{y}J_1 \,. \tag{176}$$

Substituting Eqs. (172) and (173) into Eq. (176), we obtain

$$L_3 = 0 \,, \tag{177}$$

which shows that the $L_3$-integral of screw dislocations is zero, and therefore, it is a conserved integral. Hence, there is no rotational moment about the $z$-axis produced by screw dislocations. The reason is that a screw dislocation possesses a $SO(2)$ (cylindrical) symmetry around the dislocation line, which is in the $z$-direction.

### 4.1.4. *Fundamental relations between $J$- and $M$-integrals*

From the $J$- and $M$-integrals (159), (160), (161) and (166), we find the following fundamental relations

$$J_1 = -2 \frac{\partial M}{\partial \bar{x}} \,, \tag{178}$$

$$J_2 = -2 \frac{\partial M}{\partial \bar{y}} \,, \tag{179}$$

$$J_r = -2 \frac{\partial M}{\partial \bar{r}} \,, \tag{180}$$

which show that the $J$- and $M$-integrals of screw dislocations are not distinct integrals. Especially, if the $M$-integral is given, then the $J_1$-, $J_2$-, and $J_r$-integrals can be easily obtained from the simple relations (178)–(180). We have obtained that the $J_1$-integral is twice the negative of the partial derivative of the $M$-integral with respect to the position $\bar{x}$, the $J_2$-integral is twice the negative of the partial derivative of the $M$-integral with respect to the position $\bar{y}$ and the $J_r$-integral is twice the negative of the partial derivative of the $M$-integral with respect to the distance $\bar{r}$.

Note that fundamental relations between the $J$-, $M$-, and $L_3$-integrals of straight (edge and screw) dislocations in the framework of incompatible elasticity theory of dislocations for isotropic materials have been recently derived by Agiasofitou and Lazar [2016]. It is rather interesting to see that the same relations hold also for the $J$- and $M$-integrals of screw dislocations in one-dimensional hexagonal quasicrystals.

## 5. $J$-, $M$-, and $L$-integrals of generalized compatible elasticity theory of quasicrystals

If the quasicrystalline material is free of dislocations, then the plastic fields are zero and we work within the framework of generalized compatible elasticity theory of quasicrystals.

### 5.1. *$J$-, $M$-, and $L$-integrals of dislocation-free quasicrystals*

Since the quasicrystal is dislocation-free, the Peach-Koehler force is zero. However, we still have the appearance of configurational forces; namely the Cherepanov force density due to the presence of body forces and the inhomogeneity force density or

Eshelby force density since the material is considered as a non-homogeneous one. Thus,

$$f_k^{\text{conf}} = f_k^{\text{C}} + f_k^{\text{inh}} \tag{181}$$

with

$$f_k^{\text{C}} = f_i^{\parallel} u_{i,k}^{\parallel} + f_A^{\perp} u_{A,k}^{\perp} \tag{182}$$

the *Cherepanov force density for compatible elasticity of quasicrystals* and

$$f_k^{\text{inh}} = \frac{1}{2} u_{i,j}^{\parallel} [\partial_k C_{ijmn}] u_{m,n}^{\parallel} + u_{i,j}^{\parallel} [\partial_k D_{ijBn}] u_{B,n}^{\perp} + \frac{1}{2} u_{A,j}^{\perp} [\partial_k E_{AjBn}] u_{B,n}^{\perp} , \tag{183}$$

the *inhomogeneity force density* or *Eshelby force density for compatible elasticity of quasicrystals*.

From Eqs. (42) and (60) for vanishing plastic fields, we obtain the **$J$-** and **$M$-**integrals for an arbitrary quasicrystal which is non-homogeneous in presence of body forces

$$J_k = \int_S P_{kj} \, dS_j = \int_V f_k^{\text{conf}} \, dV = \int_V \left( f_k^{\text{C}} + f_k^{\text{inh}} \right) dV , \tag{184}$$

$$M = \int_S Y_j \, dS_j = \int_V \left( x_j f_j^{\text{conf}} + \frac{d-2}{2} \left( u_j^{\parallel} f_j^{\parallel} + u_A^{\perp} f_A^{\perp} \right) \right) dV . \tag{185}$$

In the above equations, $P_{kj}$ is the Eshelby stress tensor for compatible elasticity of quasicrystals given by Eq. (46) and $Y_j$ is the dilatation or scaling flux vector given by Eq. (54) with $P_{kj}$ given by Eq. (46).

From Eqs. (83), (109) and (141), we obtain the **$L$-**integrals for one-dimensional quasicrystals, three-dimensional cubic quasicrystals and two-dimensional decagonal quasicrystals of Laue class 14 for a non-homogeneous material in presence of body forces, respectively

$$L_k = \int_S M_{kj} \, dS_j = \int_V \epsilon_{kji} \Big( x_j f_i^{\text{conf}} - u_j^{\parallel} f_i^{\parallel} + \big[ u_{j,l}^{\parallel} \sigma_{il}^{\parallel} + u_{l,j}^{\parallel} \sigma_{li}^{\parallel} + u_{3,j}^{\perp} \sigma_{3i}^{\perp} \big] \Big) dV , \tag{186}$$

$$L_k = \int_S M_{kj} \, dS_j = \int_V \epsilon_{kji} \Big( x_j f_i^{\text{conf}} - u_j^{\parallel} f_i^{\parallel} - u_j^{\perp} f_i^{\perp} + \big[ u_{j,l}^{\parallel} \sigma_{il}^{\parallel} + u_{l,j}^{\parallel} \sigma_{li}^{\parallel} + u_{j,l}^{\perp} \sigma_{il}^{\perp} + u_{l,j}^{\perp} \sigma_{li}^{\perp} \big] \Big) dV , \tag{187}$$

$$L_3 = \int_S M_{3j} \, dS_j = \int_V \Big( \epsilon_{3ji} (x_j f_i^{\text{conf}} - u_j^{\parallel} f_i^{\parallel}) - 3\epsilon_{3BA} u_B^{\perp} f_A^{\perp} \Big) dV . \tag{188}$$

In Eqs. (186) and (187), $M_{kj}$ is the angular momentum tensor given by Eqs. (74) and (100) for one-dimensional quasicrystals and for three-dimensional cubic quasicrystals, respectively. The angular momentum vector $M_{3j}$ for two-dimensional decagonal quasicrystals of Laue class 14 is given by Eq. (115).

It can be seen in the above equations that the **$J$-**, **$M$-**, and **$L$-**integrals for a non-homogeneous material in presence of body forces in the framework of generalized compatible elasticity theory of quasicrystals are not conserved integrals, since they do not vanish due to the existence of the configurational forces. In the $L_k$-integrals (186) and (187) there is an additional contribution due to the material anisotropy.

## 5.2. *J-, M-, and L-integrals of dislocation-free, homogeneous quasicrystals with vanishing body forces*

We consider here dislocation-free, homogeneous quasicrystals with vanishing body forces. In this case, the configurational or material force densities are zero. Only in this case, we have the opportunity to obtain conservation laws. From Eqs. (184) and (185), we obtain the *J- and M-integrals for an arbitrary quasicrystal, which is homogeneous in absence of body forces*

$$J_k = \int_S P_{kj} \, \mathrm{d}S_j = 0 \,, \tag{189}$$

$$M = \int_S Y_j \, \mathrm{d}S_j = 0 \,. \tag{190}$$

From Eqs. (186)–(188), we obtain the *L-integrals for one-dimensional quasicrystals, three-dimensional cubic quasicrystals and two-dimensional decagonal quasicrystals of Laue class 14 for a homogeneous material in absence of body forces*, respectively

$$L_k = \int_S M_{kj} \, \mathrm{d}S_j = \int_V \epsilon_{kji} \big( u_{j,l}^\| \sigma_{il}^\| + u_{l,j}^\| \sigma_{li}^\| + u_{3,j}^\perp \sigma_{3i}^\perp \big) \, \mathrm{d}V \,, \tag{191}$$

$$L_k = \int_S M_{kj} \, \mathrm{d}S_j = \int_V \epsilon_{kji} \big( u_{j,l}^\| \sigma_{il}^\| + u_{l,j}^\| \sigma_{li}^\| + u_{j,l}^\perp \sigma_{il}^\perp + u_{l,j}^\perp \sigma_{li}^\perp \big) \, \mathrm{d}V \,, \tag{192}$$

$$L_3 = \int_S M_{3j} \, \mathrm{d}S_j = 0 \,. \tag{193}$$

It can be seen in the above equations that the *J-* and *M*-integrals are conserved integrals. However, the $L_k$-integrals (191) and (192) are not conserved due to the material anisotropy. Only in the case that the material is isotropic, we can obtain a rotational conservation law, and consequently a conserved *L*-integral.

## 6. Conclusions

In this work, the generalization of the Eshelbian or configurational mechanics towards quasicrystals has been presented. The global translational, rotational, and dilatational (scaling) balance laws for incompatible elasticity of quasicrystals have been derived. Special attention has been payed to the investigation of the rotational transformations, since there exist many kinds of symmetries of the material coefficients for quasicrystals. Three representative examples have been examined giving rise to the exploration of the corresponding isotropy conditions. Particular interest has been given to the derivation of the *J-*, *M-*, and *L*-integrals of dislocations (dislocation loops and straight dislocations) in quasicrystals. As a concrete example, the explicit forms of the *J-*, *M-*, and *L*-integrals have been derived for the case of two parallel screw dislocations in one-dimensional hexagonal quasicrystals. It is rather interesting to see that results that hold originally for the $\boldsymbol{J}$-, $M$-, and $\boldsymbol{L}$-integrals of dislocations in the framework of incompatible elasticity theory of dislocations for isotropic materials (see [Agiasofitou and Lazar, 2016]) are also true for the $\boldsymbol{J}$-, $M$-, and $\boldsymbol{L}$-integrals of dislocations in quasicrystals. The $\boldsymbol{J}$-, $M$-, and $\boldsymbol{L}$-integrals can be of great importance for further applications in dislocation dynamics (DD), fracture mechanics as well as for dislocation based fracture mechanics of quasicrystals.

### Acknowledgements

The authors gratefully acknowledge grants from the Deutsche Forschungsgemeinschaft (Grant Nos. La1974/2-2, La1974/3-1, La1974/3-2).

### References

Agiasofitou, E., Lazar, M. and Kirchner, H. [2010] "Generalized dynamics of moving dislocations in quasicrystals," *J. Phys.: Condens. Matter* **22**, 495401 (8pp).

Agiasofitou, E. and Lazar, M. [2014] "On the equations of motion of dislocations in quasicrystals," *Mech. Res. Commun.* **57**, 27–33.

Agiasofitou, E. and Lazar, M. [2016] "Micromechanics of dislocations in solids: $\boldsymbol{J}$-, $M$- and $\boldsymbol{L}$-integrals and their fundamental relations," submitted for publication.

Bak, P. [1985] "Symmetry, stability, and elastic properties of icosahedral incommensurate crystals," *Phys. Rev. B* **32**, 5764–5772.

Bilby, B. A. and Eshelby, J. D. [1968] "Dislocations and the Theory of Fracture," In: *Fracture, An Advanced Treatise*, Liebowitz, H,. (Ed.), Academic Press, New York, pp. 100–182; Reprinted in: Collected Works of J.D. Eshelby, Markenscoff, X., Gupta, A., (Eds.) pp. 861–902, Springer, Dordrecht (2006).

Bloom, P. D., Baikerikar, K. G., Anderegg, J. W. and Sheares, V. V. [2003] "Fabrication and wear resistance of Al-Cu-Fe quasicrystal-epoxy composite materials," *Mater. Eng. A* **360**, 46–57.

Budiansky, B. and Rice, J. R. [1973] "Conservation laws and energy-release rates," *J. Appl. Mech.* **40**, 201–203.

Cherepanov, G. P. [1979] *Mechanics of Brittle Fracture* (McGraw-Hill, New York).

Cherepanov, G. P. [1981] "Invariant $\Gamma$ integrals," *Eng. Fract. Mech.* **14**, 39–58.

deWit, R. [1973] "Theory of disclinations II," *J. Res. Nat. Bur. Stand. (U.S.)* **77A**, 49–100.

Ding, D. H., Wang, W., Hu, C. and Yang, R. [1993] "Generalized elasticity theory of quasicrystals," *Phys. Rev. B* **48**, 7003–7010.

Ding, D. H., Wang, W., Yang, R. and Hu, C. [1995] "General expressions for the elastic displacement fields induced by dislocations in quasicrystals," *J. Phys.: Condens. Matter* **7**, 5423–5436.

36  *REFERENCES*

Dubois, J. M., Kang, S. S. and Von Stebut, J. J. [1991] "Quasicrystalline low-friction coatings," *J. Mater. Sci. Lett.* **10**, 537–541.

Edagawa, K. and Takeuchi, S. [2007] "Elasticity, Dislocations and their Motion in Quasicrystals," in: Nabarro, F.R.N., Hirth, J.P. (Eds.), *Dislocations in Solids 13*, Elsevier B.V., North-Holland, Amsterdam, pp. 365–417.

Eshelby, J. D. [1975] "The elastic energy-momentum tensor," J. Elast. **5**, 321–335; Reprinted in *Collected Works of J.D. Eshelby*, eds. X. Markenscoff and A. Gupta, pp. 753–767, Springer, Dordrecht (2006).

Fan, T. Y. and Mai, Y. W. [2004] "Elasticity theory, fracture mechanics, and some relevant thermal properties of quasi-crystalline materials," *Appl. Mech. Rev.* **57**, 325–343.

Fan, T. Y. [2011] *Mathematical Theory of Elasticity of Quasicrystals and Its Applications* (Science Press, Beijing and Springer-Verlag, Berlin, Heidelberg).

Feuerbacher, M. [2012] "Dislocations in icosahedral quasicrystals," *Chem. Soc. Rev.* **41**, 6745–6759.

Gel'fand, I. M. and Shilov, G. E. [1964] *Generalized Functions, Vol. I* (Academic, New York).

Gong, P., Hu, C.-Z., Zhou, X., Miao, L. and Wang, X. [2006a] "Isotropic and anisotropic properties of quasicrystals," *Eur. Phys. J. B* **52**, 477–481.

Gong, P., Hu, C.-Z., Zhou, X., Miao, L. and Wang, X. [2006b] "Nonlinear elasticities of octagonal and dodecagonal quasicrystals," *Phys. Lett. A* **356**, 168–173.

Hermann, C. [1934] "Tensoren und Kristallsymmetrie," *Z. Kristallogr.* **89**, 32–48.

Hirth, J. P. and Lothe, J. [1982] *Theory of Dislocations* 2nd edition (John Wiley, New York).

Hu, C., Wang, R., Yang W. and Ding, D. [1996] "Point groups and elastic properties of two-dimensional quasicrystals," *Acta Cryst. A* **52**, 251–256.

Hu, C., Wang, R. and Ding, D. H. [2000] "Symmetry groups, physical property tensors, elasticity and dislocations in quasicrystals," *Rep. Prog. Phys.* **63**, 1–39.

Kenzari, S., Bonina, D., Dubois, J. M. and Fournée, V. [2012] "Quasicrystal-polymer composites for selective laser sintering technology," *Mater. Design* **35**, 691–695.

Kienzler, R. and Herrmann, G. [2000] *Mechanics in Material Space* (Springer, Berlin).

Kirchner, H. [1999] "The force on an elastic singularity in a non-homogeneous medium," *J. Mech. Phys. Solids* **47**, 993–998.

Lardner, R. W. [1974] *Mathematical Theory of Dislocations and Fracture* (University of Toronto Press, Toronto).

Lazar, M. and Agiasofitou, E. [2014] "Fundamentals in generalized elasticity and dislocation theory of quasicrystals: Green tensor, dislocation key-formulas and dislocation loops," *Phil. Mag.* **94**, 4080–4101.

Lazar, M. and Kirchner, H. O. K. [2007a] "The Eshelby stress tensor, angular

momentum tensor and dilatation flux in gradient elasticity," *Int. J. Solids Struct.* **44**, 2477–2486.

Lazar, M. and Kirchner, H. O. K. [2007b] "The Eshelby stress tensor, angular momentum tensor and dilatation flux in micropolar elasticity," *Int. J. Solids Struct.* **44**, 4613–4620.

Lei, J., Wang, R., Hu, C. and Ding, D.-H. [1999] "Diffuse scattering from decagonal quasicrystals," *Phys. Rev. B* **59**, 822–828.

Levine, D., Lubensky, T. C., Ostlund, S., Ramaswamy, S., Steinhardt, P. J. and Toner, J. [1985] "Elasticity and dislocations in pentagonal and icosahedral quasicrystals," *Phys. Rev. Lett.* **54**, 1520–1523.

Li, X. Y. [2014] "Fundamental solutions of penny-shaped and half-infinite plane cracks embedded in an infinite space of one-dimensional hexagonal quasi-crystal under thermal loading," *Proc. R. Soc. A* **469**, 20130023.

Li, X. F. and Fan, T. Y. [1999] "A straight dislocation in one-dimensional hexagonal quasicrystals," *phys. stat. sol. (b)* **212**, 19–26.

Li, S. and Wang, G. [2008] *Introduction to Micromechanics and Nanomechanics* (World Scientific, Singapore).

Markenscoff, X. and Gupta, A. [2006] *Collected Works of J. D. Eshelby, The Mechanics of Defects and Inhomogeneities* (Springer, Dordrecht).

Maugin, G. A. [1993] *Material Inhomogeneities in Elasticity* (Chapman and Hall, London).

Mura, T. [1987] *Micromechanics of Defects in Solids* (Dordrecht, Martinus Nijhoff).

Park, J. Y., Ogletree, D. F., Salmeron, M., Jenks, C. J. and Thiel, P. A. [2004] "Friction and adhesion properties of clean and oxidized Al-Ni-Co decagonal quasicrystals: a UHV atomic force microscopy/scanning tunneling microscopy study," *Tribol. Lett.* **17**, 629–636.

Rice, J. R. [1968] "A path independent integral and the approximate analysis of strain concentration by notches and cracks," *J. Appl. Mech.* **35**, 379–386.

Ripamonti, C. [1987] "Physical symmetry of quasicrystals," *J. Phys. France* **48**, 893–895.

Shechtman, D., Blech, I., Gratias, D. and Cahn, J. W. [1984] "Metallic phase with long-range orientational order and no translational symmetry," *Phys. Rev. Lett.* **53**, 1951–1953.

Shi, W. [2005] "Conservation laws of a decagonal quasicrystal in elastodynamics," *Eur. J. Mech. A/Solids* **24**, 217–226.

Shi, W. [2007] "Conservation integrals of any quasicrystal and application," *Int. J. Fract.* **144**, 61–64.

Socolar, J. E. S. [1989] "Simple octagonal and dodecagonal quasicrystals," *Phys. Rev. B* **39**, 10519–10551.

Teodosiu, C. [1982] *Elastic Models of Crystal Defects* (Springer, Berlin).

Wang, R., Yang, W., Hu, C. and Ding, D.-H. [1997] "Point and space groups and elastic behaviours of one-dimensional quasicrystals," *J. Phys.: Condens. Mat-*

*ter* **9**, 2411–2422.

Wang, R. and Hu, C. [2002] "Dislocations in Quasicrystals," in: *Intermetallic Compounds - Principles and Practice: Progress, Volume 3*, Eds. J.H. Westbrook and R.L. Fleischer, John Wiley & Sons, Ltd, Chichester, UK, pp. 379–402.

Yang, W., Wang, R., Ding, D. H. and Hu, C. [1993] "Linear elasticity theory of cubic quasicrystals," *Phys. Rev. B* **48**, 6999–7003.

Zhou, X., Hu, C., Gong, P. and Qiu, S. [2004] "Nonlinear elastic properties of decagonal quasicrystals," *Phys. Rev. B* **70**, 094202 (5 pages).